\documentclass[aps,prd,longbibliography,twocolumn,amsmath,amssymb,floatfix,nofootinbib,noeprint]{revtex4-2}

\usepackage{graphicx}
\usepackage{dcolumn}
\usepackage{bm}
\usepackage{latexsym}
\usepackage{slashed}
\usepackage[colorlinks=true,linkcolor=blue,citecolor=blue]{hyperref}

\newcommand{\Rco}{$\mathcal R_{\slashed E}$}

\begin{document}

\title{Particle acceleration in strong MHD turbulence}

\author{Martin Lemoine} 
\affiliation{Institut d'Astrophysique de Paris,\\
CNRS -- Sorbonne Universit\'e, \\
98 bis boulevard Arago, F-75014 Paris, France}

\date{\today}

\begin{abstract}  
Nonthermal acceleration of particles in magnetohydrodynamic (MHD) turbulence plays a central role in a wide variety of astrophysical sites. This physics is addressed here in the context of a strong turbulence, composed of coherent structures rather than waves, beyond the realm of quasilinear theory. The present description tracks the momentum of the particle through a sequence of frames in which the electric field vanishes, in the spirit of the original Fermi scenario. It connects the sources of energy gain (or loss) to the gradients of the velocity of the magnetic field lines, in particular the acceleration and the shear of their velocity flow projected along the field line direction, as well as their compression in the transverse plane. Those velocity gradients are subject to strong intermittency: they are spatially localized and their strengths obey power law distributions, as demonstrated through direct measurements in the incompressible MHD simulation of the Johns Hopkins University database. This intermittency impacts the acceleration process in a significant way, which opens up prospects for a rich phenomenology. In particular, the momentum distribution, which is here captured through an analytical random walk model, displays extended power law tails with soft-to-hard evolution in time, in general agreement with recent kinetic numerical simulations. Extensions to this description and possible avenues of exploration are discussed. 
\end{abstract}

%\pacs{}
\maketitle

\section{Introduction}
The remarkable wealth of physical phenomena that inhabit the cascades of magnetohydrodynamic (MHD) turbulence~\cite{Beresnyak:2034323,2020arXiv201000699S,Matthaeus21}, from the large scales of stirring motions down to the microscopic dissipative layers, the ubiquity of magnetized turbulence in space plasmas, and the decisive roles that it plays in various astrophysical settings~\cite{1998RvMP...70....1B,2004ARA&A..42..211E,2016JPlPh..82f5302C,2019MNRAS.484.4881M}, have turned its study into a field of research of its own right, with broad interdisciplinary connections. In highly conducting, collisionless astrophysical plasmas, MHD turbulence indeed governs the transport of particles, including cosmic rays, it regulates the thermodynamical properties of plasmas, and of more direct concern to the present study, it promotes particle acceleration, thereby opening a connection between the fundamentals of plasma turbulence and nonthermal astronomy. 

In his seminal works~\cite{1949PhRv...75.1169F,1954ApJ...119....1F}, E. Fermi described particle acceleration in highly conducting, turbulent plasmas through the kinematics of discrete scattering events: a bounce of the particle on a region of high magnetic field strength in motion, or a turn-around of the trajectory in a curved magnetic field configuration, which, in a more modern context of guiding center theories of particle motion, can be pictured as  ``grad-B drift'' and ``curvature drift'' energization.

Most of the literature in this field nevertheless hinges on the notion of wave particle resonance in the context of quasilinear theory~\cite{66Kennel,67Hall,68Lerche,1971ApSS..12..302K} and its more recent nonlinear extensions~\cite{1973Ap&SS..25..471V,2003ApJ...590L..53M,2008ApJ...673..942Y,2009ASSL..362.....S}. In this picture, the turbulent plasma is described as a linear superposition of uncorrelated waves, {\it e.g.} Alfv\'en and magnetosonic modes in ideal magnetohydrodynamics (MHD). Energy gain then derives from phase locking of the particle trajectory with those waves. A truly remarkable aspect of quasilinear theory is its ability to provide explicit calculations of the acceleration process, up to the very shape of the accelerated spectrum, from a limited number of parameters, most notably the turbulence strength and the power spectrum of magnetic fluctuations~\cite{1984A&A...136..227S,1991A&A...250..266A,92Jaekel,98Schlick,2002cra..book.....S,2008ApJ...681.1725S,2011ApJ...732...96S}. This has triggered a wealth of applications in astrophysics, on all scales, from the Sun~\cite{96Larosa,2004ApJ...610..550P,2004MNRAS.354..870S,2008ApJ...684.1461Y,2012ApJ...754..103B} to the remote Universe~\cite{04Liu,2007MNRAS.378..245B,2008ApJ...682..175P,2011MNRAS.410..127B,2016MNRAS.458.2584B,17Eckert}, including high-energy sources harboring relativistic plasmas in extreme environments~\cite{1996ApJ...461L..37B,2011ApJ...739...66T,2016JPlPh..82d6301L,2016PhRvD..94b3005A,2017ApJ...846L..28X,18Asano,2019ApJ...872...10X}.

However, the realization~\cite{2000PhRvL..85.4656C,2002PhRvL..89B1102Y,2004A&A...420..799S} that wave particle resonances disappear in the context of modern anisotropic MHD turbulence theories~\cite{95GS,1997ApJ...485..680G,2000ApJ...539..273C,01Maron,03Cho,2006PhRvL..96k5002B,2011PhRvL.106g5001B} has cast shadows on this program, and more generally on the role of waves in turbulent dissipation. Accounting for nonlinear effects, such as resonance broadening related to the finite lifetime of the eigenmodes, can at best lead to scattering timescales of the order of the coherence scale of the turbulence, at least for for those modes that are subject to anisotropy. Consequently, this renders hazardous the extrapolation of predictions based on quasilinear calculations to the regime of strong turbulence\footnote{By ``strong turbulence'', it is meant here the large-amplitude regime in which the magnetic fluctuations are comparable to the mean field, $\delta B\gtrsim B_0$. It also implies that, all along the cascade, the turbulence will be strong in the sense defined by Goldreich and Sridhar, {\it i.e.} governed by critical balance~\cite{95GS}.}; this point will be detailed in the following Sec.~\ref{sec:waves}.

Meanwhile, {\it in situ} measurements in the solar wind~\cite{2009PhRvL.103g5006K,2011ApJ...727L..11O,2012PhRvL.108z1102O,2019E&SS....6..656B} as well as numerical simulations~\cite{2012PhRvL.109s5001W,2013PhPl...20a2303K,2014PhRvL.112u5002O} have revealed that dissipation is closely associated to coherent structures with sharp gradients, {\it e.g.} current sheets and flux tubes. It has become apparent that MHD turbulence must be described as a collection of structures rather than linear uncorrelated waves, or possibly as a combination of the two~\cite{2019PhRvX...9c1037G,2021NewA...8301507O}. 

Recent numerical simulations suggest that particles can draw energy from their interaction with the coherent structures of a turbulent bath~\cite{2018JPlPh..84f7201P,2020ApJ...894..136T}, and it is known that particles exploring a sheared velocity field with a finite mean free path can gain energy in a nonresonant fashion~\cite{1983ICRC....9..313B,1988SvAL...14..255P,1990ZhETF..98.1255B,2003ApJ...595..195W,2004ApJ...603...23C,2010ApJ...713..475J,2013ApJ...767L..16O,2014ApJ...797...28Z,2015ApJ...801..112L,2016ApJ...819...90B,2016MNRAS.458.2584B,2017ApJ...846L..28X,2019PhRvD..99h3006L}. However, a definite model able to relate the physics of particle acceleration with the characteristics of the structures of a turbulent bath remains lacking.

The aim of this paper is to make progress along those lines, in particular to study the physics of particle acceleration in strong turbulence, through nonresonant interactions with random velocity structures. To do so, it relies on a formalism that connects the sources of energy gain/loss to the shear of the $\boldsymbol{E}\times\boldsymbol{B}$ velocity field. That formalism was introduced in Ref.~\cite{2019PhRvD..99h3006L}, which considered the case of a particle exploring in an isotropic manner a random flow characterized by a single length scale. In a turbulent context, this can represent a particle whose gyroradius $r_{\rm g}$ is comparable or greater than the coherence scale $\ell_{\rm c}$ of the turbulent bath. By contrast, the present paper focuses on particles whose gyroradius is much smaller than this length scale $\ell_{\rm c}$, {\it i.e.} deep in the inertial range. It thus pays attention to the anisotropy imposed by the magnetic field line, to the approximate gyromotion of the particle around that field line, and it considers the influence of all modes of scales larger than $r_{\rm g}$.

As will be made explicit, such nonresonant acceleration has important consequences which distinguishes it from linear wave particle interactions. Notably, particle energization can be seen as a form of shear acceleration; hence energization scales with the gradients of the magnetic energy density, rather than the turbulent magnetic energy density. More importantly, the physics of acceleration becomes strongly affected by intermittency because the shear of the velocity field peaks on small scales. Different particles thus experience vastly different histories, which gives rise to power law tails of the momentum distribution function. This effect will be made manifest through an analytical random walk model that captures the effect of an inhomogeneous repartition of structures. 

While most of the discussion is analytical in nature, contact will be made with large-scale simulations of turbulence for specific points. The statistics of the random velocity structures, in particular, will be extracted from the forced MHD simulation of the Johns Hopkins Turbulence database\footnote{available from:\\ 
\href{http://turbulence.pha.jhu.edu/Forced_MHD_turbulence.aspx}
{http://turbulence.pha.jhu.edu/Forced\_MHD\_turbulence.aspx.}}, a $1024^3$ simulation of driven, incompressible MHD simulation, hereafter referred to as the JHU-MHD simulation~\cite{2008JTurb...9...31L,2013Natur.497..466E}. 

The discussion is ordered as follows. Section~\ref{sec:waves} offers a more in-depth perspective on the difference between resonant wave particle interactions and nonresonant acceleration in a random velocity flow. Section~\ref{sec:structures} sets up and describes the present model of nonresonant acceleration in a bath of strong turbulence. Section~\ref{sec:energization} provides a quantitative assessment of particle energization in the present framework; in particular, it evaluates the rates of mean and diffusive energy gains, and it calculates the spectral shape in the frame of a microscopic picture that considers the influence of turbulence intermittency. A summary and conclusions are drawn in Sec.~\ref{sec:conclusion}.

\section{Motivations: wave particle interactions in strong turbulence}\label{sec:waves}
Whether collisionless MHD turbulence should be described as a bath of waves, or structures, is a long-standing debate, see~\cite{2019PhRvX...9c1037G,2021NewA...8301507O} for recent discussions and comparison to simulation and observational data. The general point of view of the present study is that, even if wave packets are present on some scales $l$ in a bath of strong turbulence, they more likely appear as random velocity structures of ill-defined polarization to particles of gyroradius $r_{\rm g}\ll l$, than as linear wave packets.  This view relies on the following arguments.

Quasilinear theory (QLT) has been devised for weak turbulence, where the fluctuations are assumed to be much smaller in magnitude than the mean quantities. It also rests on the premise that the turbulent bath can be described as the sum of noninteracting linear waves, and that the particle collects their influence as it propagates unperturbed along the mean field~\cite{66Kennel,67Hall,68Lerche}. It has since been enlarged and generalized to account for interactions with wave packets of finite temporal coherence, and for deviations of the particle trajectory from its unperturbed orbit, in the frame of so-called extended quasilinear calculations or nonlinear guiding center theories~\cite{1973Ap&SS..25..471V,2003ApJ...590L..53M,2008ApJ...673..942Y,2009ASSL..362.....S}.

The extrapolation of quasilinear theory to the realm of large-amplitude turbulence -- $\delta B \sim B$ -- faces essential difficulties, however. A first one is that one can no longer regard the waves as uncorrelated and non-interacting, which invalidates the random phase approximation central to QLT calculations. This is notably illustrated in Ref.~\cite{01Maron} and for what regards acceleration, in Ref.~\cite{2006ApJ...637..322A}. A second one is related to critical balance~\cite{2000PhRvL..85.4656C,2002PhRvL..89B1102Y}: eddy anisotropy implies that the interaction between waves and particles becomes a nonresonant phenomenon\footnote{Fast magnetosonic modes seem to preserve an isotropic cascade and thus their efficiency in scattering particles through gyro-resonant and transit-time damping resonances~\cite{2002PhRvL..89B1102Y,03Cho,2006ApJ...638..811C}. If these modes survive in large-amplitude turbulence, they may play an important role in particle scattering and acceleration. Note that transit-time damping is in itself a form of nonresonant acceleration~\cite{19Teraki}, and as such, it will be captured in the nonresonant formalism described in this paper.}, to such a degree that all scales $\gtrsim r_{\rm g}$ contribute equally to the diffusion of particles (see below). However, the mean field direction on a given scale does not coincide with that seen from another scale, because in strong turbulence, this mean field is defined through spatial coarse-graining, scale by scale~\cite{2000ApJ...539..273C}. Consequently, the polarization of a wave on some scale $l_1$ becomes lost if viewed from another smaller scale $l_2\ll l_1$, which thus invalidates the predictions of QLT as extrapolated to the regime of strong turbulence. This point is made more explicit in the paragraphs that follow.

To understand why all scales contribute to particle energization in the quasilinear view of anisotropic turbulence, consider the QLT predictions for the momentum diffusion coefficient, 
\begin{equation}
 D_{pp}\,\sim\, \beta_{\rm A}^2\,e^2\,\int \mathrm{d} k_\perp \mathrm{d} k_\parallel \, k_\perp  \mathcal{R}_{\boldsymbol{k}} \frac{J_{1}(z_\perp)^2}{z_\perp^2}\,\mathcal S_{\boldsymbol{k}}\,,
\label{eq:Dpp_QLT}
\end{equation}
where only the dominant term has been retained, and numerical prefactors of the order of unity have been ignored; $\beta_{\rm A}\equiv v_{\rm A}/c$, $v_{\rm A}$ the Alfv\'en velocity. A detailed expression for the above can be found in, {\it e.g.} \cite{2020PhRvD.102b3003D}. The wave numbers $k_\perp$ and $k_\parallel$ are understood to be oriented perpendicular and parallel to the (scale-dependent) mean field, while the function $\mathcal S_{\boldsymbol{k}}$ represents the power spectrum. For anisotropic turbulence~\cite{95GS,1997ApJ...485..680G,2000ApJ...539..273C,01Maron,03Cho,2006PhRvL..96k5002B,2011PhRvL.106g5001B}, $\mathcal S_{\boldsymbol{k}}\propto \langle\delta B^2\rangle k_\perp^{-10/3}k_{\rm min}^{1/3}\,g\left[k_\parallel/\left(k_\perp^{2/3}k_{\rm min}^{1/3}\right)\right]$, where $g(x)$ is a function peaked around 0 and spread around $[-1,+1]$, of integral unity, which characterizes the anisotropy. $J_1(z_\perp)$ is a Bessel function, with $z_\perp = k_\perp r_{\rm g\perp}$, $r_{\rm g\perp}$ the particle Larmor radius. 

The quantity of interest, here, is the resonance function $\mathcal{R}_{\boldsymbol{k}}$, which characterizes the strength/duration of the interaction between waves and particles. Accounting for wave damping at a rate $\vert\Im\omega\vert \sim k_\parallel v_{\rm A}$ broadens the resonance and confers to $\mathcal{R}_{\boldsymbol{k}}$ a Lorentzian or Gaussian shape. Performing the integration over $k_\parallel$ then brings the integral under the form 
\begin{equation}
D_{pp}\propto  p^2 \,k_{\rm min} c\,\int_\rho^{+\infty}{\rm d}z_\perp z_\perp^{-3} J_1\left(z_\perp\right)^2\,,
\label{eq:Dpp_QLT_2}
\end{equation}
and $\rho\equiv k_{\rm min}r_{\rm g}$ ($k_{\rm min}$ the minimum wave number of the cascade) is assumed to be much smaller than unity, for particles interacting with modes in the inertial range of the turbulent spectrum. Most of the contribution comes from the interval $\rho\leq z_\perp\leq 1$, corresponding to spatial scales larger than $r_{\rm g}$, yet smaller than $\ell_{\rm c}\equiv 2\pi/k_{\rm min}$, the integral scale of the turbulent spectrum. The small argument limit of the Bessel function implies that the integrand scales as $z_\perp^{-1}$. This confirms the above claim that, in this QLT description of anisotropic turbulence, all modes of scale larger than $r_{\rm g}$ contribute equally to the scattering of the particle.

Now, the anisotropy of the cascade constrains the angle $\theta_k$ between the direction of the longitudinal axis of an eddy and that of the mean field to $\theta_k\lesssim k_\parallel/k$. On small scales, meaning large $k$, $k_\parallel \propto k_\perp^{2/3}$ implies $\theta_k\sim (k_\perp/k_{\rm min})^{-1/3}$. However, on scale $l \sim 1/k\sim 1/k_\perp$, the field direction itself undergoes angular excursions of magnitude $\delta \theta\sim\delta B_l/B \sim \eta^{1/2}\left(k_\perp/k_{\rm min}\right)^{-1/3}$, where $\eta = \langle\delta B^2\rangle/B^2$ denotes the ratio of the turbulent magnetic energy density to the total magnetic energy density and $\delta B_l^2 \sim k\langle\delta B_k^2\rangle\vert_{k\sim 1/l}\sim k_\perp^2k_\parallel S_{\boldsymbol{k}}\vert_{k\sim 1/l}$. Consequently, in large-amplitude turbulence, $\eta\sim1$, those excursions are, on all scales, comparable to $\theta_k$ and the wave polarization is ill-defined. As anticipated above, this invalidates the QLT calculation, which rests on Eq.~(\ref{eq:Dpp_QLT}).

The above compares the directions of the magnetic field after coarse-graining on two different scales, but at the same physical point, whereas the particle travels along the line as it cumulates the effect of waves. Given the intrinsic stochasticity of magnetic field lines in strong turbulence~\cite{2013Natur.497..466E}, the correlation between the direction of the magnetic field, scale to scale, is bound to be lost even more rapidly. Small fluctuations in the magnetic field line direction can be incorporated within the frame of extended quasilinear theories as an additional resonance broadening attached to the randomization of the particle pitch angle, see {\it e.g.}~\cite{1975RvGSP..13..547V,18Xu,2020PhRvD.102b3003D}, but this does not change the above conclusions.

Yet another implication of strong turbulence is that a wave packet on a given scale $l_1$ is nonlinearly distorted on a timescale comparable to its wave period~\cite{01Maron}, hence during the interaction itself it loses its polarization and characteristics of linear waves. 

In summary, therefore, from the point of view of a particle propagating in a strongly turbulent environment, large-scale wave packets -- assuming that they represent a fair representation of the turbulence -- are more likely seen as imprinting velocity and magnetic fluctuations, rather than as waves contributing to wave particle interactions that are anyhow nonresonant given their inherent anisotropy. 

Numerical simulations provide essential tools to test such models, but their interpretation deserves special care. In Ref.~\cite{14Lynn}, for instance, it was shown that the above predictions of (extended) quasilinear theory, {\it viz.} a momentum diffusion coefficient $D_{pp} \sim v_{\rm A}^2 \, p^2\,/(\ell_{\rm c}c)$, could account for the energization of test-particles in MHD simulations of mildly sub-Alfv\'enic incompressible turbulence. However, that agreement cannot be interpreted as a confirmation of quasilinear predictions because the scaling of that diffusion coefficient is, as a matter of fact, generic to any nonresonant acceleration mechanism. The diffusion coefficient indeed scales, for dimensional reasons, as $D_{pp}\propto p^2/t_{\rm int}$, where $t_{\rm int}$ defines some interaction timescale. In the absence of resonant interactions, the integral length scale $\ell_{\rm c}$ sets this interaction time, {\it viz.} $t_{\rm int} \sim \ell_{\rm c}/v$ (with $v$  the particle velocity), while actual resonances would introduce another scale in the problem ($r_{\rm g}$) and lead to the estimate $t_{\rm int} \propto r_{\rm g}^{2-q}$, with $q$ the index of the power spectrum of magnetic fluctuations~\cite{2002cra..book.....S}. The nonresonant acceleration mechanisms discussed thereafter, which do not involve waves but velocity structures, will also yield estimates that match those numerical simulations.

Finally, recent kinetic numerical simulations have contradicted, in a rather direct way, the generic predictions of a quasilinear picture of a homogeneous bath of waves: Refs.~\cite{2020ApJ...894..136T,2020arXiv201203043N} have observed that particles can be accelerated in localized regions, while Refs.~\cite{17Zhdankin,2018ApJ...867L..18Z,2018MNRAS.474.2514Z,18Comisso,2020ApJ...893L...7W,2019ApJ...886..122C,2019PhRvL.122e5101Z,2019arXiv190606306N} have demonstrated that the particle distribution function exhibit non-thermal power law tails, in stark contrast with the standard lognormal spectrum predicted by QLT. Such power law spectra appear to be related to the intermittent nature of the acceleration process~\cite{2020MNRAS.499.4972L}. The prevalent role of intermittency and the development of power law spectra will emerge as natural consequences of the nonresonant acceleration scenario that is discussed next.

\section{Particle acceleration in random velocity structures}\label{sec:structures}
\subsection{General framework}
Within the frame of ideal MHD, a particle can gain energy and diffuse in momentum space as it explores the shear of the motional electric field in the plasma. It is the shear, rather than the magnitude of the electric field that matters, because in any region devoid of shear, one can boost to a frame in which this electric field vanishes, and there the particle undergoes helicoidal motion around the magnetic field, at constant energy.

To describe such random interactions, it proves convenient to define the frame \Rco~in which this motional electric field vanishes. With respect to the lab-frame, \Rco~moves at velocity (three-velocity, measured in units of $c$) $\boldsymbol{\beta_E} \equiv \boldsymbol{E}\times \boldsymbol{B}/B^2$ (Gaussian units are adopted all throughout). More specifically, following Ref.~\cite{2019PhRvD..99h3006L}, the idea is to track the motion of the particle in physical space in the lab-frame, and the history of its momentum in this \Rco~frame. This provides a generalized picture of the classical Fermi mechanism, which is usually described as elastic interactions in the frame of pointlike, discrete magnetized scattering centers. The present picture is well suited to describe particle transport in a continuous random flow of velocity structures.

An important difference with respect to quasilinear theory is that the calculation for energy gain is performed in this frame \Rco, not in the lab frame. For subrelativistic turbulence, in which the rms velocity fluctuation $\langle\delta u_{\rm p}^2\rangle^{1/2}\ll c$, this amounts to a simple velocity shift. Nevertheless, it will be shown in the following that the statistics of the field line velocity field $\boldsymbol{u_E}$ differ from those of the plasma velocity. 

The formalism that follows is fully relativistic because relativity offers the requisite tools to describe the history of particles through this continuous sequence of (noninertial) reference frames. This relativistic description is furthermore mandatory to make contact with kinetic numerical simulations of relativistic turbulence, which have revealed broad momentum spectra of accelerated particles~\cite{17Zhdankin,2018ApJ...867L..18Z,2018MNRAS.474.2514Z,18Comisso,2020ApJ...893L...7W,2019ApJ...886..122C,2019PhRvL.122e5101Z,2019arXiv190606306N}, and which thus offer ideal experiments for the present model. Subrelativistic approximations will nevertheless be provided where opportune.

As $\boldsymbol{\beta_E}$ is a random quantity, the \Rco~frame is not globally inertial, but the electric field is zero everywhere along the particle trajectory. Consequently, energy gains and losses are entirely expressed as inertial forces felt by the particle as the \Rco~frame accelerates or decelerates. One clear advantage of this point of view is to relate in a direct way the statistics of energy gain to those of the velocity field. 

A key ingredient is how the particle sees, or travels through, the velocity structures. While Ref.~\cite{2019PhRvD..99h3006L} assumed that the particle explored in an isotropic manner a random velocity field whose variations were characterized by a single length scale, the present work considers the more realistic case of a particle subject to a crossed electric - magnetic field configuration with perturbations spanning a broad range of scales. The former study thus applies to particles whose gyroradius is comparable or larger than the coherence scale $\ell_{\rm c}$, while the present generalizes it to particles in the inertial range, that is $r_{\rm g}\ll\ell_{\rm c}$. Those particles are sensitive to fluctuations on all scales larger than their gyroradius as they gyrate around the local mean field. On the other hand, fluctuations on scales smaller than $r_{\rm g}$ can be neglected here, as commonly done, because they carry a small fraction of the power if $r_{\rm g}\ll\ell_{\rm c}$ and because their influence averages out over a gyroperiod. 

The acceleration of particles is thus described here as a continuous sequence of interactions with moving magnetized structures on scales larger than the gyroradius. This implies that all quantities -- velocity flow, magnetic field in particular -- must be understood as coarse-grained over the scale $r_{\rm g}$. Formally, this means that we consider a velocity field 
${\boldsymbol{\beta_E}}_l = {\boldsymbol{E}}_l\times{\boldsymbol{B}}_l/B_l^2$, where ${\boldsymbol{E}}_l$ and ${\boldsymbol{B}}_l$ are as before defined through coarse-graining on scale $l$, which differs from the ``true'' $\boldsymbol{\beta_E}$ by $\mathcal O\left(\delta B_l^2/B^2\right)$. Note that the field entering the ``true'' contribution is obtained without coarse-graining here, although this limit is somewhat ambiguous in the frame of an MHD description, which breaks down on kinetic scales (see~\cite{2018PhRvX...8d1020E}, in particular). Let us stress at this point that ideal MHD is not a requisite of the present approach, which only needs to assume that parallel electric fields play a negligible role in particle acceleration. This means, in particular, that the electric field characterizing the velocity field $\boldsymbol{\beta_E}$ can incorporate other terms of Ohm's law beyond the ideal one, such as, for instance, the Hall MHD term. 

For the sake of clarity and simplicity, all indices $_l$ referring to coarse-graining will be suppressed when not necessary. All quantities expressed in the frame \Rco~will be primed. Furthermore, all four-velocities $u'$ in \Rco~are from now on expressed in units of $c$, so that $u'=p'/(mc)$ in terms of the momentum $p'$, and similarly for the four-velocity of the magnetic field lines, $u_E=\beta_E/\sqrt{1-{u_E}^2}$ (unless otherwise noted).

Given the above scheme of approximation, the \Rco~frame is mainly characterized by two  four-vectors: ${u_E}^\alpha$, the (timelike) four-velocity of \Rco, and $b^\alpha$, the unit spacelike four-vector indicating the direction of the magnetic field line. This construction is made more explicit in Appendix~\ref{sec:appA}, which also provides the details of the calculations. 

In \Rco, the particle Lorentz factor $\gamma'$ and four-velocity component parallel to the mean field, $u_\parallel'$, are then found to evolve as follows:
\begin{align}
\frac{1}{c}\frac{{\rm d}\gamma'}{{\rm d}\tau}&\,=\,-{u'}^b{u'}^c\,{{\mathsf e}^\beta}_b{{\mathsf e}^\gamma}_c\,\frac{\partial}{\partial x^\gamma} {u_E}_\beta   \,,\nonumber\\
\frac{1}{c}\frac{{\rm d}u_\parallel'}{{\rm d}\tau}&\,=\, -{u'}^b{u'}^c\,{{\mathsf e}^\beta}_b{{\mathsf e}^\gamma}_c\,\frac{\partial}{\partial x^\gamma} {b}_\beta \,.
\label{eq:Rco-evol-0}
\end{align}
In this equation, ${u'}^b$, with $b\in\left\{0,\dots,\,3\right\}$, represents a component of the particle four-velocity and ${{\mathsf e}^\beta}_b$ is a component of the vierbein that connects lab-frame quantities to quantities evaluated in \Rco, {\it i.e.} which provides the instantaneous Lorentz transform between those two frames. Note, in particular, that the spatiotemporal derivative $\partial_\gamma\,\equiv\,\partial/\partial x^\gamma$ in the above equation is expressed in lab-frame quantities. On the other hand, the derivative on the lhs is taken with respect to proper time $\tau$ which parametrizes the particle trajectory. The connection between the statistics of the velocity flow of the magnetic field lines and particle energization is manifest in Eq.~(\ref{eq:Rco-evol-0}).

On scales larger than $r_{\rm g}$, we can assume that, at each point along the trajectory, the particle  undergoes a helicoidal orbit around the (curved) magnetic field line. In this case, the above equation can be further simplified, see Appendix~\ref{sec:appA},
\begin{align}
\frac{1}{c}\frac{{\rm d}\gamma'}{{\rm d}\tau}&\,=\,-\gamma'u_\parallel'\,\boldsymbol{a_E}\cdot\boldsymbol{b} - {u_\parallel'}^2\,\Theta_\parallel- \frac{1}{2}{u_\perp'}^2\Theta_\perp\,,\nonumber\\
\frac{1}{c}\frac{{\rm d}u_\parallel'}{{\rm d}\tau}&\,=\,-{\gamma'}^2\, \,\boldsymbol{a_E}\cdot\boldsymbol{b} - \gamma'u_\parallel'\,\Theta_\parallel - \,\frac{1}{2}{u_{\perp}'}^2\boldsymbol{b}\cdot\boldsymbol{\nabla}\ln B'\,,
\label{eq:Rco-evol}
\end{align}
introducing the components of the shear of the field line four-velocity, parallel ($\Theta_\parallel$) and perpendicular ($\Theta_\perp$) to the magnetic field line, as well as its acceleration $\boldsymbol{a_E}$, \begin{align}
\Theta_{\parallel}&\,\equiv\,b^\alpha b^\beta \partial_\alpha {u_E}_\beta\,,\nonumber\\
\Theta_{\perp}&\,\equiv\,\left(\eta^{\alpha\beta}  - b^\alpha b^\beta\right) \partial_\alpha {u_E}_\beta\,,\nonumber\\
{a_E}^\alpha&\,\equiv\,{u_E}^\beta\partial_\beta {u_E}^\alpha\,.
\label{eq:thetaperp-redef}
\end{align}

\begin{figure}
\includegraphics[width=0.48\textwidth]{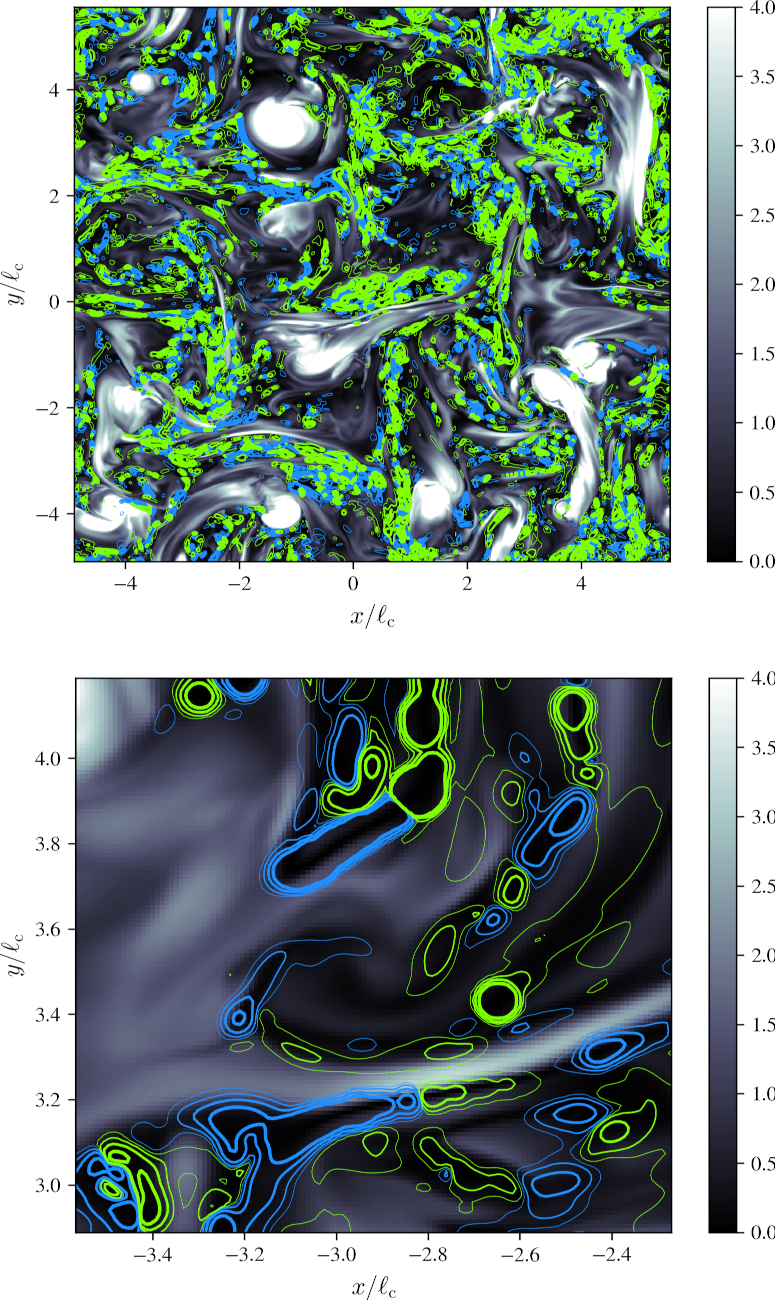}
\caption{Density plot of the magnetic energy density (black and white, in units of the average energy density) in a two-dimensional slice of the JHU-MHD simulation cube, with overlayed contours of $\Theta_\perp^2$ at (from thin to thick, in blue, solid lines): $0.5,\,2.,\,8.,\,32.$ in units of $\left(v_{\rm A}/\ell_{\rm c}\right)^2$, and similarly for negative values in yellow. Top panel: $1024^2$ resolution, extending over about $10\ell_{\rm c}\times10\ell_{\rm c}$; bottom panel: zoom over a region of $1.3\ell_{\rm c}\times 1.3\ell_{\rm c}$.
\label{fig:Bmap}}
\end{figure}

The first line of Eq.~(\ref{eq:Rco-evol}) represents the main equation governing the acceleration of particles in random velocity structures in our current scheme of approximation. The force contributions that enter this equation find a natural interpretation. The first term $\propto \boldsymbol{a_E}\cdot\boldsymbol{b}$ describes the effective gravity that the particle feels in the \Rco~frame in the direction of the magnetic field line as this line accelerates or decelerates. The second term $\propto \Theta_\parallel$ captures energy gain or loss through shear acceleration along the direction of motion of the particle (along the field line). Similarly, the particle can gain (or loose) energy if the field lines, to which it is attached, are compressed (or dilated) in their transverse plane; this gives rise to the third contribution.

In the sub-relativistic regime, $\beta_E\ll1$, $u_E\simeq \beta_E$, and the above expressions can be approximated as 
\begin{align}
&\Theta_\parallel\simeq \boldsymbol{b}\cdot\left(\boldsymbol{b}\cdot\boldsymbol{\nabla}\right)\boldsymbol{u_E}\,\nonumber\\
&\Theta_\perp\simeq\boldsymbol{\nabla}\cdot\boldsymbol{u_E}-\Theta_\parallel\,\nonumber\\
&\boldsymbol{a_E}\simeq(1/c)\partial_t\boldsymbol{u_E}\,.
\label{eq:approx-subrel}
\end{align}
The acceleration is thus one order higher in $\beta_E$ than the shear components because of the time derivative; consequently, its contribution can be neglected when $\beta_E\ll1$. 

In the present scheme, no particular difference arises between particles of different energies. Although the rhs of Eq.~(\ref{eq:Rco-evol}) scales as the square of the particle four-velocity in \Rco, one power is absorbed by the fact that the derivative is taken with respect to proper time $\tau$ on the lhs, because ${\rm d}\tau = {\rm d}t/\gamma$ in the lab frame (here, $\gamma$ denotes the particle Lorentz in the lab frame). The remaining power implies that energy gain/loss scales in proportion to the particle energy, as in the original Fermi model. 

Figure~\ref{fig:Bmap} provides a contour map of the quantity $\Theta_\perp^2$ in the JHU-MHD simulation of incompressible turbulence, superimposed on a density plot of the magnetic field energy density. The distribution of $\Theta_\parallel^2$ is not shown here, but comparable to that of $\Theta_\perp^2$. The contribution of the gravitational term $\propto \boldsymbol{a_E}\cdot\boldsymbol{b}$ is  smaller than that of the other two; for the JHU-MHD turbulence simulation, $\langle {u_E}^2\rangle^{1/2} \simeq 0.41$ (units of $c$). Note that, for linear Alfv\'en waves, $\Theta_\perp=\Theta_\parallel=0$ exactly, while for linear magnetosonic modes, $\Theta_\parallel=0$ but $\Theta_\perp\sim \beta_{\rm MS}\,k\,\delta B/B\neq0$, where $\beta_{\rm MS}$ denotes the phase velocity of magnetosonic modes (units of $c$), $k$ the wave number and $\delta B/B$ the relative mode amplitude.

In the context of quasilinear theory, particle acceleration scales with the power spectrum of magnetic fluctuations, hence with the turbulent magnetic energy density, while in the present nonresonant description, acceleration scales with the shear of the field line velocity flow. Figure~\ref{fig:Bmap} thus provides a vivid illustration of the difference between these two pictures. It shows in particular that $\Theta_\perp^2$ tends to take large values in regions of strong gradients of the magnetic energy density, with some preference for regions with lower than average magnetic energy density. Such regions are subject to the magnetic pressure from the neighbouring regions of strong $B$; at the same time, small scale modes exert a larger (relative) influence on the magnetic field direction in zones of magnetic depressions, which can also induce strong gradients.

This correlation with the gradients of the fields is of particular importance in the context of modern turbulence theories, as it directly connects particle acceleration with turbulence intermittency, and with the coherent structures that characterize this intermittency.

\subsection{Connection to plasma heating/cooling}
The two terms $\Theta_\perp$ and $\Theta_\parallel$ appear in different combinations in Eq.~(\ref{eq:Rco-evol}), because of the nature of particle gyromotion, but they generalize the idea that a particle trapped in a flow undergoing compression or dilation gains or looses energy. Assume indeed, for the time being, that the particle four-velocity is isotropic at all times: the acceleration contribution then vanishes because $\langle u_\parallel'\rangle=0$, while both longitudinal and transverse shear contributions add up into the isotropic combination
\begin{equation}
\frac{1}{c}\frac{{\rm d}\gamma'}{{\rm d}\tau}\,=\,-\frac{1}{3}{u'}^2\,\partial_\alpha {u_E}^\alpha\,, 
\label{eq:isotropic-contrib}
\end{equation}
as in~\cite{2019PhRvD..99h3006L} for isotropic scattering. In the subrelativistic limit, $\partial_\alpha{u_E}^\alpha\simeq \boldsymbol{\nabla}\cdot\boldsymbol{\beta_E}$. 

In a fluid description, adiabatic heating/cooling through compression/dilation is expressed through a similar operator, although it is expressed in terms of the fluid velocity $\boldsymbol{u_{\rm p}}$, not the velocity of field lines $\boldsymbol{u_E}$. The fluid approximation tacitly assumes that particles are isotropized in the fluid rest frame through some effective collision, while in the collisionless case, what governs the energization is the velocity of the field lines to which the particle is attached.

The quantities $\boldsymbol{\nabla}\cdot\boldsymbol{u_E}$ and $\boldsymbol{\nabla}\cdot\boldsymbol{u_{\rm p}}$ differ from one another. To express this difference and relate the gradients of $\boldsymbol{u_E}$ to the sources of heating and cooling, we assume here $u_E\ll1$ to make contact with subrelativistic MHD; similar developments can be found in \cite{2015ApJ...801..112L,2018ApJ...855...80L}. Note first that
\begin{equation}
\boldsymbol{u_E} = \boldsymbol{u_{\rm p}} - \boldsymbol{b}\left(\boldsymbol{u_{\rm p}}\cdot\boldsymbol{b}\right)\,,
\label{eq:uE-to-uf}
\end{equation}
so that 
\begin{equation}
\boldsymbol{\nabla}\cdot\boldsymbol{u_E}\,=\,\boldsymbol{\nabla_\perp}\cdot\boldsymbol{u_{\rm p}} + \frac{\boldsymbol{u_{\rm p}}\cdot\boldsymbol{B}}{B^4}\boldsymbol{B}\cdot\boldsymbol\nabla B^2
-\frac{1}{B^2}\boldsymbol{u_{\rm p}}\cdot\left(\boldsymbol{B}\cdot\boldsymbol{\nabla}\right)\boldsymbol{B}\,,
\label{eq:uE-to-uf-2}
\end{equation}
which indicates that the difference between the compression of $\boldsymbol{u_E}$ and that of $\boldsymbol{u_{\rm p}}$ mainly relates to field line curvature. 

From Poynting's theorem to first order in the electric field contribution,
\begin{equation}
\frac{1}{2c}\frac{\partial}{\partial t} B^2\,=\,-\boldsymbol{E}\cdot\left(\boldsymbol{\nabla}\times\boldsymbol{B}\right) - \boldsymbol{\nabla}\cdot\left(\boldsymbol{E}\times\boldsymbol{B}\right)\,,
\label{eq:sub-rel-Pth}
\end{equation}
we obtain
\begin{equation}
\boldsymbol{\nabla}\cdot\boldsymbol{u_E}\,=\,-\frac{1}{2B^2 c}\frac{{\rm d}}{{\rm d}t}B^2 - \frac{1}{B^2}\boldsymbol{u_E}\cdot\left(\boldsymbol{B}\cdot\boldsymbol{\nabla}\right)\boldsymbol{B}\,,
\label{eq:uE-to-uf-3}
\end{equation}
with ${\rm d}/{\rm d}t = \partial_t + c\boldsymbol{u_E}\cdot\boldsymbol{\nabla}$. This equation relates the compression of the field lines to the change of magnetic energy density along the field line trajectory and to the field line tension. The last term on the rhs of the above equation, in particular, can be recast as $\Theta_\parallel$, given that $\boldsymbol{u_E}\cdot\boldsymbol{B}=0$. Therefore, 
\begin{equation}
\Theta_\perp\,=\,\boldsymbol{\nabla_\perp}\cdot\boldsymbol{u_E}\,=\,-\frac{1}{2B^2 c}\frac{{\rm d}}{{\rm d}t}B^2  \,.
\label{eq:uE-to-uf-4}
\end{equation}
This relation can also be obtained directly from the advection equation. In the fully relativistic regime, its analog is provided in Appendix~\ref{sec:appA},
\begin{equation}
\frac{1}{2{B'}^2}\frac{1}{c}\frac{\partial}{\partial t'}{B'}^2 \,=\,-\Theta_\perp\,,
\label{eq:uE-to-uf-5}
\end{equation}
providing a direct connection between the tranverse compression of field lines and the temporal evolution of the magnetic energy density in the \Rco~frame.

The terms that contribute to energization, up to the inertial one, are thus connected to the sinks to magnetic energy. In this regard, the acceleration process which is depicted here does not differ much from the physics of heating in a plasma of infinite conductivity and, to a first approximation, those particles that get accelerated to high energies can be regarded as those that have had the chance to encounter more zones of heating than the rest of the population.  The shape of the spectrum at high-energy is then determined by sequences of outlier scattering events. Those will be captured in Sec.~\ref{sec:e-interm} through the use of large deviation theory.

\subsection{Connection to the guiding center description}
Equation~(\ref{eq:Rco-evol}) has been obtained under the assumption that the particle experiences inertial forces in the \Rco~frame, on spatiotemporal scales larger than its gyration around the magnetic field line, with proper averaging over this gyromotion and in the frame of ideal MHD. For this reason, it captures to this order the contribution of guiding center theories (at their first order).

The detailed relationships connecting the terms of Eq.~(\ref{eq:Rco-evol}) to those of guiding center theories are provided in Appendix~\ref{sec:appA}. It is shown, in particular, that the term involving $\Theta_\parallel$ represents the curvature drift, while the term $\Theta_\perp$ captures the grad-B drift, {\it i.e} the mirror force responsible for transit-time damping acceleration, and the gravitational term is connected to the inertial drift. Equation~(\ref{eq:uE-to-uf-5}) above also  expresses the betatron nature of the grad-B drift. 

These two schemes of approximation remain of a different nature, however, and differences should be seen at higher orders. Theories of guiding center motion are perturbative schemes based on expansions in $r_{\rm g}/L$, where $L\sim \ell_{\rm c}$ represents the typical length scale on which the electromagnetic fields evolve. In the framework of Hamiltonian guiding center models, one defines at each order an adiabatic invariant conjugate to a gyrofrequency, its invariance expressing the fact that the short timescales of gyromotion have been averaged out of the dynamics. Such theories are difficult to expand, however, given their inherent sophistication, {\it e.g.}~\cite{1987PhDT.......197B,1999PhPl....6.4487B,2009RvMP...81..693C}, and the equations of motions, written for quantities that are redefined order by order, rapidly take a complicated form.

By contrast, Eq.~(\ref{eq:Rco-evol-0}) is exact, as it simply expresses the equation of motion in a frame in which the electric field vanishes. The equation that we use in this paper, Eq.~(\ref{eq:Rco-evol}), is itself an approximation of that equation, to lowest order in $r_{\rm g}/\ell_{\rm c}$: namely, the derivatives $\partial_\gamma {u_E}_\beta$ and $\partial_\gamma b_\beta$ have been calculated at the position of the guiding center and the perpendicular velocity components ${u'_2}$ and ${u'_3}$ have been averaged out over this gyromotion. To go to higher orders, one would nevertheless start from Eq.~(\ref{eq:Rco-evol-0}) again and include different approximations for those terms. 

In this sense, the present formalism may allow for a more phenomenological approach to the problem than guiding center theories, which are based on a specific hierarchy in $r_{\rm g}/\ell_{\rm c}$. For instance, Sec.~\ref{sec:energization} will account for particle diffusion along the field line, so that the particle position in Eq.~(\ref{eq:Rco-evol}) will be treated as a random variable, in the spirit of extensions to quasilinear theories~\cite{1973Ap&SS..25..471V,2003ApJ...590L..53M,2008ApJ...673..942Y,2009ASSL..362.....S}. A similar improvement over the present description would be to account for perpendicular transport in space, through field-line wandering combined with particle jump from field line to field line. 

One should also consider the influence of drifts beyond the classical $\boldsymbol{E}\times\boldsymbol{B}$ drift: such terms, spurred by large-scale modes, would modify the lowest order form for the particle four-velocity ${u'}^b$ and therefore generate additional force terms, of higher order in the power of the gradients of the velocity field. As mentioned earlier, one may also consider introducing terms beyond ideal Ohm's law in the definition of the electric field, to capture physics at smaller length scales. Regarding small-scale modes, one extension that deserves investigation, is their contribution to the velocity of the field line, as alluded to earlier. This effect, of order $\mathcal O\left(\delta B_l^2/B^2\right)$, would also introduce higher powers of the velocity fluctuations.

\subsection{Violation of the adiabatic invariant and pitch-angle scattering}\label{sec:deltaM}
The adiabatic invariant $M\,\equiv\, {u_\perp'}^2/(2B')$ is conserved to the present order of approximation, as can be explicitly verified using Eq.~(\ref{eq:Rco-evol}). Such conservation represents an obvious barrier to stochastic acceleration to large momenta, because $B'$ can be regarded as a bounded quantity, just as the pitch angle cosine in \Rco, $\mu'$. In this regard, higher order effects, such as those discussed above, are a necessity; see also Refs.~\cite{2018JPlPh..84f7201P,2021arXiv210209654C} for a related discussion. An important issue, therefore, is the hierarchy between the rate at which $M$ is violated and that at which particles can gain momentum according to the model described by Eq.~(\ref{eq:Rco-evol}). We argue here that those two rates are likely of the same order of magnitude. 

Note first that the inertial forces that control energization also contribute to scattering, as in any Fermi process. In particular, Eq.~(\ref{eq:Rco-evol}) implies
\begin{align}
\frac{1}{c}\frac{{\rm d}\mu'}{{\rm d}\tau}\,=\,&
-\frac{{\gamma'}^2}{u'}\left(1-{\mu'}^2\right)\boldsymbol{a_E}\cdot\boldsymbol{b}
-\gamma'\mu'\left(1-{\mu'}^2\right)\Theta_\parallel\nonumber\\
&\,\,-\frac{1}{2}u'\left(1-{\mu'}^2\right)\boldsymbol{b}\cdot\boldsymbol{\nabla}\,\ln B'
\nonumber\\
&\,\,+\frac{1}{2}\gamma'\,\mu'\,\left(1-{\mu'}^2\right)\Theta_\perp\,.
\label{eq:Rco-scatt}
\end{align}
In a turbulence of small r.m.s. velocity fluctuation, the third term, which corresponds to the mirror force, dominates, because all others scale with $u_E$. This suggests that particle scattering would be dominated in that regime by magnetic mirroring effects. In strong turbulence, such mirrors may differ from their linear magnetosonic mode counterparts, as discussed in Sec.~\ref{sec:waves}. However, what matters to particle acceleration, here, is chaoticity, and in that respect, one must distinguish the violation of $M$, which at fixed $u'$ and $B'$ corresponds to a stochastic excursion of $\mu'$, from the regular, nonchaotic evolution of the pitch-angle cosine in a $M-$conserving environment.

Interactions that violate the conservation of $M$ likely arise from small-scale perturbations and/or from regions of magnetic depressions where a magnetic fluctuation $\delta B'$ can strongly distort the mean field direction. At one extreme, large-scale modes ($l\gg r_{\rm g}$) renormalize the adiabatic invariant by contributions of order $\delta M/M \sim \mathcal O\left[\left(r_{\rm g}/l\right)\left(\delta B'/B'\right)\right]$~\cite{1981PhFl...24.1730L}. At the other extreme, small scale perturbations ($l\ll r_{\rm g}$) impart a deflection
$\delta u_\perp'/u_\perp' \sim \mathcal O\left[\left(l/r_{\rm g}\right)\left(\delta B'/B'\right)\right]$ over their length scale $l$. This brings a change in $M$ to the order $\delta M/M = 2\delta u_\perp'/u_\perp' - \delta B'/B'$. 

At a fixed value of $\delta B'/B'$, the comparison of the above large-scale and small-scale influences suggest that the dominant contribution comes from scales comparable to the gyroradius. However,  intermittency implies that large excursions in $\delta B'/B'$ become more frequent at smaller length scales, which means that, overall, order unity violations of $M$ are more likely to occur from small scale modes. Shock waves or reconnection layers provide examples of microscopic phenomena that can lead to substantial variation in $\delta B'$.

Magnetic perturbations $\delta B'/B'$, when coarse-grained on spatial scales $\gtrsim r_{\rm g}$, also set the magnitude of the ($M-$conserving) forces that control pitch-angle scattering, see Eq.~(\ref{eq:Rco-scatt}), hence acceleration, given that $\Delta u'/u' \sim \mu'\Delta \mu'/(1-\mu'^2)$ at fixed $M$ and $B'$. Consequently, one can reasonably expect that the rate of $M-$violating events is comparable -- or even larger, if one takes into account the contributions of modes $l< r_{\rm g}$ that can violate $M$ but do not contribute to energization in the present model -- to that of acceleration. Finally, acceleration itself takes place on timescales larger than the gyrotime $t_{\rm g}$ because it results from the influence of modes on scales $l>r_{\rm g}$; consequently, approximate gyromotion remains a valid approximation at all times. This validates the assumptions that underlie the main Eq.~(\ref{eq:Rco-evol}). 

Another way to picture the energization process is thus the following. At fixed $M$, the phase space variables $u'$ and $\mu'$ evolve in a banana-shaped island defined by ${u'}^2(1-{\mu'}^2)=2 M B'$, whose width is controlled by the excursions of $B'$ around its mean. Islands corresponding to two neighboring values of $M$ are therefore not disconnected, and the junction of all islands for all values of $M$ span the whole phase space. Hence, provided that the rate of evolution of $M$ is comparable or larger to that of $u'$, acceleration takes place at the rate fixed by Eq.~(\ref{eq:Rco-evol}).

\section{Particle energization}\label{sec:energization}
The objective of this section is to quantify particle energization in the frame of the present model. It is split into three main parts. The first Sec.~\ref{sec:e-drift} calculates the one- and two-point statistics of energy gain, that is the mean energy gain $\left\langle\Delta p'\right\rangle$ and the diffusive term $\left\langle{\Delta p'}^2\right\rangle$. Emphasis is placed here on the ambiguity that arises in the calculation of the latter quantity, given that the force terms that control energization display highly non-Gaussian features characteristic of intermittency. Section~\ref{sec:e-interm} then studies the acceleration process in a microscopic picture, which describes the momentum trajectory as a discrete random walk in a set of intermittent structures. This will clarify the role of intermittency and explain how it shapes the momentum distribution of accelerated particles. Finally, Sec.~\ref{sec:e-disc} provides some closing remarks.

\subsection{Drift vs diffusion}\label{sec:e-drift}
\subsubsection{Mean drift}
In a purely kinematic description of Fermi acceleration, particles gain energy through head-on collisions with moving magnetized scattering centers, lose energy through tail-on collisions. On average, particles suffer more head-on than tail-on collisions, hence Fermi acceleration comes with a net (log-)momentum gain $\langle \Delta \ln p\rangle/\Delta t$ (in the laboratory frame); the latter is comparable, in order of magnitude, to the stochastic diffusive term $\langle \Delta \ln^2 p\rangle/\Delta t$. 

The present model departs from the Fermi scenario in one fundamental way: as the particle interacts with a region in which $\Theta_\parallel\neq0$ or $\Theta_\perp\neq0$, it does not gain or loose energy according to whether the interaction is head-on or tail-on, but according to whether $\Theta_\parallel$ (or $\Theta_\perp$) is negative or positive; see Eq.~(\ref{eq:Rco-evol})\footnote{The inertial term $\boldsymbol{a_E}\cdot\boldsymbol{b}$ also depends on the sign of the particle momentum; we neglect this (generally subdominant) term here, for the sake of the discussion.}. Inside any such region, the acceleration process is regular, akin to a Fermi type-1 process, and it is rather through its encounter with many different sites of alternating polarity that the particle undergoes a random walk in momentum space. Consequently, the particle population gains (or loose) energy at a net rate depending on the sign of $\langle \Theta_\parallel\rangle$ and $\langle \Theta_\perp\rangle$, where the average is computed over the spatial volume (assuming a homogeneous density distribution of particles). 

As discussed earlier, those mean values relate to turbulent plasma heating (in the MHD approximation) through large scale motions. In the JHU-MHD turbulence simulation, $\langle\Theta_\parallel\rangle \simeq +0.03 v_{\rm A}/\ell_{\rm c}$, while $\langle\Theta_\perp\rangle \simeq -0.1 v_{\rm A}/\ell_{\rm c}$, and $\langle\boldsymbol{a_E}\cdot\boldsymbol{b}\rangle\simeq 0$.
Assuming a homogeneous and isotropic distribution of particles, this implies net heating at a rate
\begin{align}
\frac{{\rm d}\langle\ln p'\rangle}{\gamma'{\rm d}\tau}&\,=\,\frac{1}{3}\left(\langle\Theta_\parallel\rangle+\langle\Theta_\perp\rangle\right)\nonumber\\
&\,\simeq\,0.02\,{\rm v_A}/\ell_{\rm c}\,.
\label{eq:heat-rate}
\end{align}
Those values do not depend on the coarse-graining spatial scale $\Delta x$ that is adopted in calculating the spatial average, provided $\Delta x < \ell_{\rm c}$. On scales $\Delta x > \ell_{\rm c}$, all averages rapidly tend to zero, which is expected insofar as energy is continuously injected on scales $\ell_{\rm c}$ in that simulation. In a steady state, energy is thus spread over the cascade, down to microscopic scales where it is dissipated through nonideal processes. In the frame of the present model, the energy contained in sheared motions on scales $l$ contributes to net heating for particles with gyroradius $r_{\rm g}\lesssim l$.  

The above heating rate indicates a rather small drift of the particle distribution on an eddy turnover timescale $\ell_{\rm c}/v_{\rm A}$. In the context of particle acceleration, what truly matters, however, is the acceleration timescale, which will be found to be of the order of $c\ell_{\rm c}/v_{\rm A}^2$ for that same simulation. Extrapolating this to other regimes suggest that, at small values of $v_{\rm A}$, heating may dominate over particle acceleration, while at larger values of $v_{\rm A}$ (relativistic range), acceleration may dominate over the drift. A word of caution is needed to specify what is understood by $v_{\rm A}$ here: in the JHU-MHD simulation, it corresponds to $\langle u_E^2\rangle^{1/2}$, meaning that it represents a rms velocity fluctuation, not a standard Alfv\'en velocity defined with respect to a coherent magnetic field. As Eq.~(\ref{eq:Rco-evol}) makes clear, indeed, what controls the heating (and the acceleration) is the magnitude of velocity fluctuations. 

Those remarks illustrate another difference of the present scheme with respect to quasilinear theory, where the velocity that enters the diffusion coefficient is the phase velocity of the linear waves, defined relatively to the mean field.

\subsubsection{Second-order moment}\label{sec:e-diff}
The diffusion coefficient can be derived as usual (in \Rco) from
\begin{equation}
D_{p'p'}\,=\,
\underset{\Delta t'\,\rightarrow\,+\infty}{\rm lim}\,\frac{1}{2\Delta t'}\int{\rm d}\tau_1{\rm d}\tau_2\,
\left\langle\frac{{\rm d}p'}{{\rm d}\tau_1}\frac{{\rm d}p'}{{\rm d}\tau_2}\right\rangle\,.
\label{eq:diff0}
\end{equation}
with $p'$ the particle momentum in \Rco. For reference, the equation of evolution of the momentum reads
\begin{equation}
\frac{1}{c}\frac{{\rm d}p'}{{\rm d}t'}\,=\,-\gamma'mc \,\mu_\parallel'\,\boldsymbol{a_E}\cdot\boldsymbol{b} - p'\,{\mu_\parallel'}^2\,\Theta_\parallel- \frac{1}{2}\,p'\,\left(1-{\mu'}^2\right)\Theta_\perp\,.
\label{eq:Rco-evol-p}
\end{equation}

The average in Eq.~(\ref{eq:diff0}) involves the correlation function at two different space-time points of the three force terms in Eq.~(\ref{eq:Rco-evol}), including their cross-correlation functions. It also involves a limit $\Delta t \rightarrow +\infty$, which, less formally, means that the time interval should be much larger than the coherence time of the random force that is exerted on the particle. This point will be discussed in greater detail further below.

The main contribution to the integral is rewritten using
\begin{equation}
\int {\rm d}t'_1{\rm d}t'_2 \left\langle Q_1\,Q_2\right\rangle
\,=\,
2\Delta t'\,\int{\rm d}k\,\mathcal R_k\,\mathcal S^Q_k\,,
\label{eq:diff0-1}
\end{equation}
with the following notations: $Q_1$ (respectively $Q_2$) denotes one of the three random forces, that is the inertial, the parallel shear or the perpendicular compressive term, calculated at the particle position at time $t_1'$ (resp. $t_2'$). On the rhs,  $\mathcal S^Q_k\,=\,\left\langle{Q_k}^2\right\rangle$ represents the (one-dimensional in $k-$space) power spectrum of $Q$. We have neglected here the cross-correlations between those terms, for the sake of clarity, and because those cross-correlations are smaller in magnitude (as checked in the JHU-MHD simulation). We have also assumed, as is customary, a random-phase approximation, which allows us to write the rhs of Eq.~(\ref{eq:diff0-1}) as a one-dimensional integral over wave numbers, thus separating the contributions mode by mode. Our goal here is not to provide a detailed calculation of the diffusion coefficient, but to illustrate the ambiguity that arises in this calculation when non-Gaussian features arise, and to motivate the model developed in the next sub-section. Those assumptions are thus not restrictive. The resonance function $\mathcal R_k$ captures as in Sec.~\ref{sec:waves} the duration of the interaction with the structure. We assume here ballistic trajectories, in which case $\mathcal R_k\sim 1/(k_\parallel v)$. Finally, we will also assume for definiteness a Goldreich-Sridhar type spectrum, in which case $k\rightarrow k_\perp$ in Eq.~(\ref{eq:diff0-1}) and $k_\parallel\rightarrow k_{\perp}^{2/3}k_{\rm min}^{1/3}$, with $k_{\rm min}\sim 1/\ell_{\rm c}$.

\begin{figure}
\includegraphics[width=0.4\textwidth]{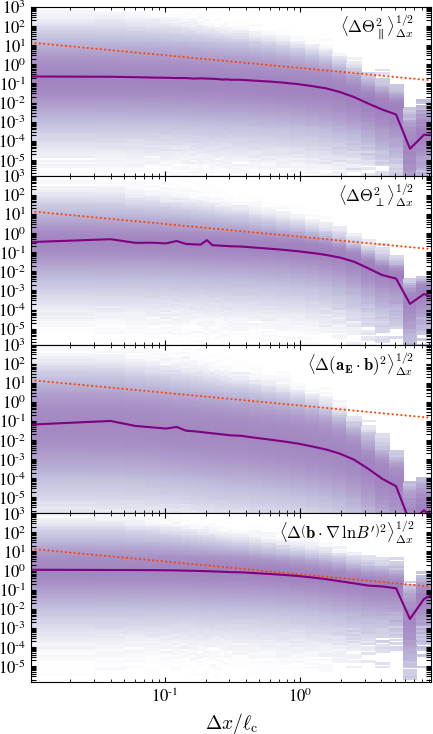}
\caption{Scaling of the r.m.s. of the terms contributing to particle energization and scattering, as indicated: inertial $\boldsymbol{a_E}\cdot\boldsymbol{b}$, parallel shear $\Theta_\parallel$ and perpendicular compression $\Theta_\perp$ (all three in units of $v_{\rm A}/\ell_{\rm c}$), and for scattering, the mirror term $\boldsymbol{b}\cdot\boldsymbol{\nabla}\ln\, B'$ in units of $1/\ell_{\rm c}$, as extracted from the JHU-MHD turbulence simulation and plotted as a function of smoothing scale $\Delta x$ (units of $\ell_{\rm c}$). The dotted line shows the scaling $\propto \Delta x^{-2/3}$; see the text for details.
\label{fig:var-theta}}
\end{figure}

Because $Q$ involves a gradient, one generally expects $S^Q_k \propto k^2 S^{u_E}_k$, with $S^{u_E}_k\propto k^{-q_u}$ the (one-dimensional in $k-$space) power spectrum of fluctuations of $u_E$. However, $\Theta_\perp$ represents a perpendicular gradient, while $\Theta_\parallel$ is a gradient measured parallel to the mean field, hence $S^{\Theta_\perp}\sim {k_\perp}^2 S^{u_E}_k$ and $S^{\Theta_\parallel}\sim {k_\parallel}^2 S^{u_E}_k$. The above calculation of the diffusion coefficient would then result in the following scalings (for $q_u=5/3$):
\begin{align}
D_{p'p'}^{\Theta_\parallel}&\,\sim\,{p'}^2\,\langle{\delta u_E}^2\rangle\,k_{\rm min}c\,\ln\left(\frac{\overline k}{k_{\rm min}}\right)\,,\nonumber\\
D_{p'p'}^{\Theta_\perp}&\,\sim\,{p'}^2\,\langle{\delta u_E}^2\rangle\,k_{\rm min}c\,\left(\frac{\overline k}{k_{\rm min}}\right)^{2/3}\,,
\label{eq:diff0-per1}
\end{align}
where $\langle{\delta u_E}^2\rangle$ provides the overall amplitude (squared) of $u_E$ velocity fluctuations.
The momentum $\overline k$ represents the maximum wave number up to which the integral can be performed. This wave number is set by the coarse-graining procedure, which considers only scales above $r_{\rm g}$, {\it i.e.} $\overline k\sim r_{\rm g}^{-1}$, or by the maximum wave number of the cascade $k_{\rm max}$, whichever is smaller. While the parallel shear contribution involves all scales above $r_{\rm g}$, the perpendicular compressive one peaks on small scales. 

The above calculation recovers the quasilinear prediction for transit-time damping in fast mode waves (described by the $\Theta_\perp$ contribution in its linear limit) if one assumes isotropic eddies: setting $k_\parallel\sim k_\perp\sim k$, one then obtains $D_{p'p'}\propto \left(\overline{k}/k_{\rm min}\right)^{1/3}$, which indeed matches the standard result~\cite{98Schlick,19Teraki}.

The ambiguity that arises in the above calculation is related to the scaling of $S^Q_k$ with $k$, once $Q$ departs from a Gaussian random field. This problem is illustrated by the statistics of the parallel shear and perpendicular compressive term extracted from the JHU-MHD turbulence simulation. Those are presented in Fig.~\ref{fig:var-theta}, expressed as $\langle Q^2\rangle_{\Delta x}^{1/2}$, meaning the r.m.s. coarse-grained over scale $\Delta x$. The correspondence with the spectra in $k-$space is obtained as usual, through $\langle\Theta_\perp^2\rangle_{\Delta x} \sim \left. k S^{\Theta_\perp}_k\right\vert_{k\sim {\Delta x}^{-1}}$ at $k\sim k_\perp$. We thus expect $\langle\Theta_\perp^2\rangle_{\Delta x}\,\propto\, {\Delta x}^{-3+q_u}\,\propto {\Delta x}^{-4/3}$ for $q_u=5/3$, but
$\langle{\Theta_\parallel}^2\rangle_{\Delta x}\,\propto\, {\Delta x}^{-7/3+q_u}\,\propto {\Delta x}^{-2/3}$ for $q_u=5/3$. 

The scaling $\langle Q^2\rangle_{\Delta x}^{1/2}\,\propto\, {\Delta x}^{-2/3}$ (expected for $\Theta_\perp$ and $\boldsymbol{a_E}\cdot\boldsymbol{b}$) is overplotted as a dotted line in Fig.~\ref{fig:var-theta}. It appears to represent fairly well the statistics of the inertial term, but not those of $\Theta_\perp$, which flatten on small scales to an apparent $\langle Q^2\rangle_{\Delta x}^{1/2}\,\propto\, {\Delta x}^0$. The parallel contribution $\Theta_\parallel$, as well as the mirror term $\langle\left(\boldsymbol{b}\cdot\boldsymbol{\nabla}\ln\,B'\right)^2\rangle_{\Delta x}^{1/2}$, also flattens out, instead of the expected $\langle Q^2\rangle_{\Delta x}^{1/2}\,\propto\, {\Delta x}^{-1/3}$. Such flat scalings would correspond to $S^Q_k \propto k^{-1}$, hence to diffusion coefficients strongly dominated by the large scales,
\begin{align}
D_{p'p'}^{\Theta_\parallel}&\,\sim\,{p'}^2\,\langle{\delta u_E}^2\rangle\,k_{\rm min}c\,,\nonumber\\
D_{p'p'}^{\Theta_\perp}&\,\sim\,{p'}^2\,\langle{\delta u_E}^2\rangle\,k_{\rm min}c\,.
\label{eq:diff0-per2}
\end{align}

\begin{figure}
\includegraphics[width=0.4\textwidth]{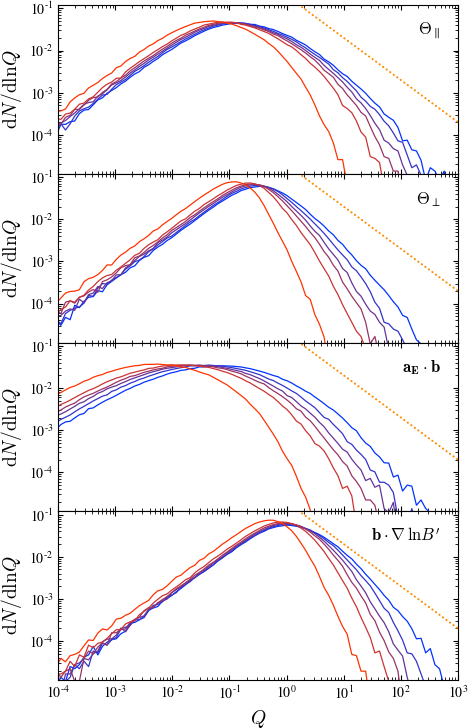}
\caption{Normalized distribution of the absolute values of the main contributions to the energization and scattering, for different smoothing scales: $\Delta x/\ell_{\rm c}=\left(0.01,\,0.09,\,0.18,\,0.26,\,0.43,\,1.1\right)$, ordered from blue to red, or equivalently right to left. From top to bottom: $\vert\Theta_\parallel\vert$; $\vert\Theta_\perp\vert$; $\vert \boldsymbol{a_E}\cdot\boldsymbol{b}\vert$; $\vert\boldsymbol{b}\cdot\boldsymbol{\nabla}\ln\,B'\vert$. The first three are expressed in units $v_{\rm A}/\ell_{\rm c}$, while the last is expressed in units $1/\ell_{\rm c}$. Each subpanel plots ${\rm d}N/{\rm d}\ln Q$, where $Q$ represents one of the above quantities. The departure from Gaussian statistics, in particular the existence of power law tails is a clear signature of intermittency. The dotted line represents a power law of index $-1$.
\label{fig:dist-theta}}
\end{figure}

Out of the two results for the diffusion coefficients, Eqs.~(\ref{eq:diff0-per1}) and (\ref{eq:diff0-per2}), the latter appears more reliable, because based on explicit measurements in the JHU-MHD simulation. Neither is however able to capture the physics of particle acceleration, for about the same reason that a second-order moment cannot suffice to characterize a non-Gaussian distribution. This is nicely illustrated in Fig.~\ref{fig:dist-theta}, which plots the probability distribution function of those force terms for various smoothing scales, as extracted from the JHU-MHD turbulence simulation. 

This figure reveals that the mean of the inertial contribution shifts to larger values as the coarse-graining scale decreases, in general agreement with the scaling observed in Fig.~\ref{fig:var-theta}, and with the expectations for a Gaussian random field. By contrast, the distribution functions ${\rm d}N/{\rm d}\ln Q$ of the three other quantities, namely $\Theta_\parallel$, $\Theta_\perp$ and $\boldsymbol{b}\cdot\boldsymbol{\nabla}\ln\,B'$, develop power law tails up to large values, with slopes close to (yet slightly larger than) $-1$. The mean of those three quantities hardly shifts as the coarse-graining scale decreases, but the slope of the power law becomes harder\footnote{Similar power laws have been recently reported for the curvature of magnetic field lines in Refs.~\cite{2019PhPl...26g2306Y,2020ApJ...898...66Y}.}. This explains, at least qualitatively, the tendency of $\langle Q^2\rangle_{\Delta x}^{1/2}$ to fall  below the predicted scaling in Fig.~\ref{fig:var-theta}.

In this context, note that the variances shown in Fig.~\ref{fig:var-theta} have been evaluated as the median of the squared fluctuations, instead of their mean, in order to avoid an excessive source of noise associated to the finite statistics given the frequent, large excursions of the quantities. Those peculiar statistics are directly related to the scarcity of the structures, quite apparent in Fig.~\ref{fig:Bmap}: the force terms can take large values in isolated regions of small spatial extent. This, in turn, directly impacts the acceleration process; in particular, it shapes the power law of accelerated particles, as discussed next.

\subsection{Random walk in intermittent structures}\label{sec:e-interm}
\subsubsection{Overview}
The classical description of turbulence intermittency comes through the fractal $\beta-$model, which ascribes to daughter eddies a probability of being turned on as the parent cascades down~\cite{1978JFM....87..719F}. Said otherwise, active daughter eddies do not span the full volume of their parent eddy, but only a filling fraction, which is characterized by a fractal Hausdorff dimension $D<3$. It is defined by $D=\ln N/\ln 2$, if each parent eddy creates $N$ daughter eddies out of $2^3$ possible ones, when cascading from scale $l$ to scale $l/2$. Correspondingly, at scale $l$, the filling fraction of active eddies is $f_l \sim f_{\ell_{\rm c}}\,(l/\ell_{\rm c})^{(3-D)}$, with $f_{\ell_{\rm c}}$ the filling fraction of active regions on the integral scale $\ell_{\rm c}$. Note that Fig.~\ref{fig:Bmap} already suggests that $f_{\ell_{\rm c}}$ takes values below unity. The value of  $D$ can take different values depending on the shape of the dissipative structures, for instance $D=1$ for filaments, $D=2$ for sheets. Its value, hence the filling fraction of active eddies, may depend on the general characteristics of the turbulence~\cite{2004PhRvL..92s1102P}.

The mean free path to interaction with random, uncorrelated structures of extent $l$ and filling fraction $f_l$ in three-dimensional space is $l_{\rm int} \sim f_l^{-1} l$ or, for $f_l$ as calculated above, $l_{\rm int} \sim f_{\ell_{\rm c}}^{-1}\ell_{\rm c}\left(l/\ell_{\rm c}\right)^{D-2}$, which can take values much larger than $l$ if $D$ takes values below 3, and even larger than $\ell_{\rm c}$ if $D\lesssim 2$. The probability of interaction per distance traveled in a fractal structure has been calculated and studied in detail in Ref.~\cite{2003PhRvE..67b6413I}. It leads to a slightly more complex form for the mean free path to interaction. Interestingly, it demonstrates that the probability of ``no interaction'' remains nonzero up to infinite distances if $D<2$. In such a case, the random walk can be modeled as a continuous time random walk with a L\'evy type distribution in waiting time. This situation has been considered in Ref.~\cite{2020MNRAS.499.4972L}, where it is shown that the momentum distribution then takes on a (near) power law shape at all times. We will ignore this possibility in the following.

The intermittent nature of acceleration sites introduces a new timescale $t_{\rm int} = l_{\rm int}/v$, whose magnitude (relative to other timescales) controls the phenomenology, if large enough. Here, $t_{\rm int}$ sets the coherence time of the random force that acts on particles. Consequently, a diffusion coefficient becomes meaningful only on time intervals $\Delta t \gg t_{\rm int}$. On timescales $\Delta t < t_{\rm int}$, different particles undergo vastly different histories, some interacting many times, others none. As discussed further below, this gives rise to the emergence of power law spectra. Such spectra have been observed recently in simulations of test-particle transport in synthetic fractal turbulent environments~\cite{2020ApJ...895L..14S}. 

On large timescales, $\Delta t \gg t_{\rm int}$, particles have had time to interact with many acceleration sites, hence one expects, in the spirit of the central limit theorem, to recover the predictions of a standard random walk in momentum space. However, this convergence will be found to be asymptotic, in the sense that the spectrum will preserve a power law tail at large momenta at all times. This will be made more explicit further below.

\subsubsection{Random walk in intermittent turbulence}
We determine the time-dependent distribution function in a microscopic picture, which describes a particle undergoing a random walk in a bath of isolated structures, of extent $l$ and filling fraction $f=f_++f_-\,\ll\,1$. Here, $f_+$ (respectively $f_-$) represents the filling fraction of those structures giving rise to energy gain $\Delta p'/p'=+g$ (respectively energy loss $\Delta p'/p'=-g$). In the frame of the present model, those structures have a negative (respectively positive) sign of $\Theta_\parallel$ or $\Theta_\perp$. We drop the primes in this section, for clarity. 

At each step of length $l$, the particle thus has a probability $f_-$ of loosing a fraction $g$ of its momentum, a probability $f_+$ of gaining $g$, and a probability $f_\emptyset=1-f_--f_+$ of crossing a void region, which corresponds to an inactive eddy in the framework of the $\beta-$model. The time-dependent solution to this random walk is captured thanks to the methods of large deviation theory~\cite{2009PhR...478....1T}, which is ideally suited to model outlier statistics before central limit convergence. The details are provided in Appendix~\ref{sec:appB}. At time $t$, this distribution can be approximated as
\begin{align}
\frac{{\rm d}N}{{\rm d}\ln p}\,\sim\,\frac{1}{g\hat t}\,&\exp\biggl\{
-\frac{\ln \hat p}{g}\ln\left[\frac{f_\emptyset\,\ln\hat p+S_p
}{2f_+\left(\ln\hat p_{\rm max}-\ln \hat p\right)}\right]\nonumber\\
& \quad\quad+\frac{\ln \hat p_{\rm max}}{g}\ln\left[\frac{f_\emptyset\,\ln\hat p_{\rm max} + 
S_p}{\ln^2\hat p_{\rm max}-\ln^2 \hat p}\right]\biggr\}\,,
\label{eq:LDPsolp}
\end{align}
with
\begin{equation}
S_p\,\equiv\,
\sqrt{f_\emptyset^2\ln^2\hat p+4f_-f_+\left(\ln^2\hat p_{\rm max}-\ln^2 \hat p\right)}\,.
\end{equation}
This makes use of the following short-hand notations: $\hat p = p/p_{\rm inj}$, where $p_{\rm inj}$ denotes the momentum at which particles are injected in the turbulent bath; $\hat t = t/\tau$, where $\tau$ denotes the time between two steps, so that, correspondingly, $\hat t$ indicates the number of steps taken; $\ln\hat p_{\rm max} = g\hat t$, which represents the maximum log-momentum that the particle can achieve in $t$. We can assume $\tau \simeq l/v$ ($v$ the particle velocity), which corresponds to ballistic trajectories. 

\begin{figure}
\includegraphics[width=0.48\textwidth]{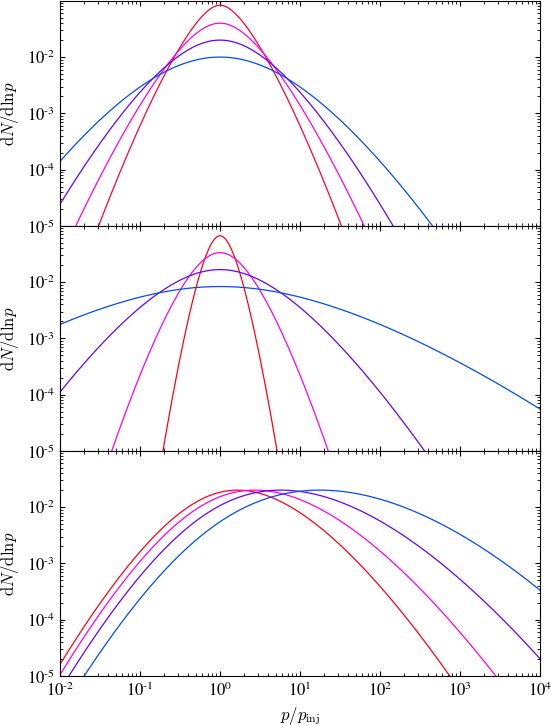}
\caption{Spectra ${\rm d}N/{\rm d}\ln p$ from the stochastic acceleration of particles in intermittent turbulence, as obtained from Eq.~(\ref{eq:LDPsolp}), see also the discussion in Appendix~\ref{sec:appB}. Top panel: spectra plotted at different (normalized) times $\hat t = 12.,\,25.,\,50.,\,100.$ (ordered from red to blue or left to right), for $g=1$ and $f_+=f_-=0.01$. Note that $\hat t=t/\tau$ characterizes the number of steps taken, hence the average number of structures encountered is $(f_++f_-)\hat t$, which takes values between $0.25$ and $2$ here. Middle panel: spectra plotted at time $\hat t=50.$, for different values of $g = 0.3,\,0.6,\,1.2,\,2.4$ (ordered from red to blue or left to right), and $f_+=f_-=0.01$. Bottom panel: spectra plotted at time $\hat t=50.$, for different values of $f_+ = 0.02,\,0.03,\,0.045,\,0.68$ (ordered from red to blue or left to right), $g=1$ and $f_-=0.01$.
\label{fig:turb-acc}}
\end{figure}

\begin{figure}
\includegraphics[width=0.48\textwidth]{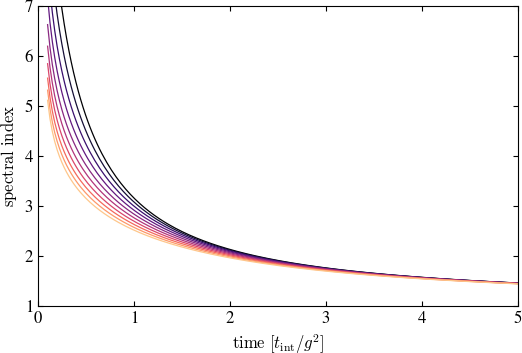}
\caption{Spectral index of the power law tail, as derived from Eq.~(\ref{eq:LDPsolp_sind}) and plotted as a function of time in units of $t_{\rm int}/g^2 = l/\left[g^2\left(f_++f_-\right)v\right]$, for $g$ spanning the range from $0.3$ to $3$ in steps of $0.3$ (from bottom to top, or dark to light). This figure assumes $f_+=f_-$. This spectral index is measured here at a momentum $p=10p_{\rm inj}$.
\label{fig:sindex}}
\end{figure}

Two combinations of the parameters play a particular role: $t_{\rm int}=\tau/\left[f_++f_--\left(f_+-f_-\right)^2\right]\sim l/(f_++f_-)v$, the interaction time with active regions, and $f_+-f_-$, which controls the amount of net heating. The momentum distribution function displays the following properties:
\begin{enumerate}
    \item The mean log-momentum $\langle \ln \hat p\rangle$ scales in proportion to $f_+-f_-$; see Fig.~\ref{fig:turb-acc} for an illustration. Defining this mean log-momentum as the peak of $p^3 f(p,t)\sim {\rm d}N/{\rm d}\ln p$, we find $\langle \ln \hat p\rangle = g \hat t\left(f_+-f_-\right)$. This result agrees with the discussion of Sec.~\ref{sec:e-drift}, where it is explained that the drift scales with the spatial average of positive minus negative contributions, that is $f_+-f_-$.

    \item The distribution possesses a log-normal core with an extended power law tail. For the log-normal core, the variance can be expressed as
    \begin{align}
    \left\langle\Delta\ln^2\hat p\right\rangle&\,=\,
    \left\langle \ln^2 \hat p - \langle\ln\hat p\rangle^2\right\rangle \nonumber\\
    &\,=\, g^2\hat t\left[f_++f_--\left(f_+-f_-\right)^2\right]\,,
    \label{eq:LDPsolp_varcore}
    \end{align}
    Consequently, at any time, the particle distribution appears to follow a diffusive process characterized by the diffusion coefficient,
    \begin{align}
    D_{pp} &\,=\, g^2\,p^2\,\frac{f_++f_--\left(f_+-f_-\right)^2}{\tau}\nonumber\\
    &\,=\, g^2\frac{p^2}{t_{\rm int}}\,,
    \label{eq:LDPsolp_diff}
    \end{align}
    introducing the effective interaction timescale $t_{\rm int} =\tau/\left[f_++f_--\left(f_+-f_-\right)^2\right]$. Note that $t_{\rm int}$ approximately matches the time between two interactions with active eddies, $l/(f_++f_-)c$, at small values of $f_+$ and $f_-$. For $t_{\rm int}\propto p^0$ and $g\propto p^0$, as discussed below, we recover the scaling $D_{pp}\propto p^2$ observed in numerical simulations. 

    \item This diffusion coefficient only represents an average value, measured over the particle population. It characterizes the general broadening $\left\langle\Delta\ln^2\hat p\right\rangle$ of the core of the spectrum, but not its shape at large momenta, which takes on a power law shape. This general spectral shape -- thermal log-normal core with power law extension -- matches that observed in recent kinetic numerical simulations~\cite{17Zhdankin,2018ApJ...867L..18Z,2018MNRAS.474.2514Z,18Comisso,2020ApJ...893L...7W,2019ApJ...886..122C,2019PhRvL.122e5101Z}. 

    \item The spectral index $s$, defined through $p^2f(p,t)\propto {\rm d}N/{\rm d}p\propto p^{-s}$, can be derived as
    \begin{align}
    s&\,=\,1 - \frac{{\rm d}}{{\rm d}\ln p}\ln\,\left[\frac{{\rm d}N}{{\rm d}\ln p}\right]\nonumber\\
    &\,=\,
    1 - \frac{1}{g}\ln\left[\frac{f_\emptyset\,\ln\hat p+S_p}{2f_+\left(\ln\hat p_{\rm max}-\ln \hat p\right)}\right]\,.
    \label{eq:LDPsolp_sind}
    \end{align}
    The power law is thus approximate, as the index runs logarithmically with $\ln\hat p$. This noteworthy result reveals that the extrapolation of the spectral index measured in numerical simulations -- which are restricted in dynamic range -- to a turbulent cascade extending over many orders of magnitude, is not straightforward. The power law tail extends here from the r.m.s. $\langle\Delta\ln \hat p^{\,2}\rangle^{1/2}\sim g \left(t/t_{\rm int}\right)^{1/2}$ to $\ln \hat p_{\rm max}=g t/\tau$ ; its extent in log-space scales as $\left(t/t_{\rm int}\right)^{1/2}/f$.

    \item The power law shape hardens with time, just as it hardens with increasing energy gain per interaction. Those features are illustrated in Fig.~\ref{fig:turb-acc}. Those properties also agree with the measurements of power law spectra in kinetic simulations of relativistic turbulence~\cite{17Zhdankin,2018ApJ...867L..18Z,2018MNRAS.474.2514Z,18Comisso,2020ApJ...893L...7W,2019ApJ...886..122C,2019PhRvL.122e5101Z}, which show that the spectrum hardens with time, with increasing magnetization and with increasing fluctuation level $\delta B/B$. Note that the energy gain per scattering event is related to these latter two quantities, magnetization and turbulence amplitude. 

    \item The spectral index mainly depends on $g$ and $t_{\rm int}$, for $f_+\sim f_-$. In the power law range, indeed, $S_p\simeq f_\emptyset\ln\hat p$ and $2f_+\left(\ln\hat p_{\rm max}-\ln\hat p\right)\sim 2f_+\ln\hat p_{\rm max}\sim g t/t_{\rm int}$. More specifically, at time $t\sim t_{\rm int}/g^2$, $s$ takes values close to $3$ for $g\sim 1$, see Fig.~\ref{fig:sindex}. The quantity $t_{\rm int}/g^2$ represents the average acceleration timescale, hence this spectral index matches nicely the values $s\sim3$ observed in kinetic simulations of relativistic turbulence if $t_{\rm int}\sim \ell_{\rm c}/c$.

\end{enumerate}

Finally, the spectrum converges asymptotically to the log-normal shape, as announced earlier. This is manifest from Eqs.~(\ref{eq:ldp-0}) and (\ref{eq:ldp-4}) in the limit $\mathcal N\rightarrow +\infty$ (with $\mathcal N=\hat t$ the number of steps taken). This property of large deviation functions derives from the central limit theorem. On asymptotic timescales, diffusion then takes place as in homogeneous turbulence because the particles can probe all structures many times over, although the effective interaction timescale is now set by $t_{\rm int}$, and the acceleration timescale by $t_{\rm int}/g^2$. To be more precise, the distribution function preserves at all times a nonthermal power law tail, which is pushed to larger and larger values of the momentum, because the Gaussian core extends farther in $\ln p$, given that $\langle\Delta\ln \hat p^{\,2}\rangle^{1/2}\sim g \left(t/t_{\rm int}\right)^{1/2}$. The convergence to the standard log-normal shape is thus asymptotic.

In kinetic numerical simulations, the power law slows down its evolution in time once the maximum energy has reached the confinement limit, {\it viz.} $r_{\rm g}\left(p_{\rm max}\right)\sim \ell_{\rm c}$, which takes place on a timescale $t \sim \left(t_{\rm int}/g^2\right)\ln\left[\ell_{\rm c}/r_{\rm g}(p_{\rm inj})\right]$~\cite{2018ApJ...867L..18Z}. Acceleration slows down at larger momenta, because the particle decouples from the turbulence (its scattering timescale increases fast with energy)~\cite{2019PhRvD..99h3006L}. This bottleneck results in a piling-up of the momenta close to this limit. It is not captured by the present random walk, which considers a momentum-independent energy gain per scattering event. However, the above timescale is significantly longer than the acceleration timescale $t_{\rm int}/g^2$, all the more so if the dynamic range of the cascade $\ell_{\rm c}/r_{\rm g}(p_{\rm inj})$ is large.

\subsubsection{Estimates for $g$ and $t_{\rm int}$}
The above analytical model assumes a given value for $g$, $t_{\rm int}$ and the scale $l$. Those can be estimated from the values that maximize the acceleration rate. Here, we extract them from the JHU-MHD simulation, using two different methods. We first consider a given coarse-graining scale $l\gtrsim r_{\rm g}$, then compare the contributions of various scales $l$. 

\begin{figure}
\includegraphics[width=0.48\textwidth]{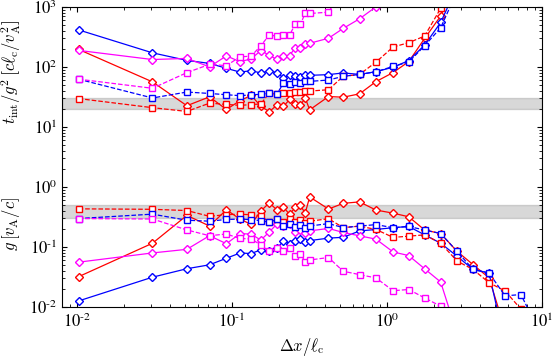}
\caption{Estimates of the effective acceleration timescale $t_{\rm int}/g^2$ (upper curves, in units of $c\ell_{\rm c}/v_{\rm A}^2$, assuming $v\simeq c$) and of the average energy gain per interaction $g$ (lower curves, in units of $v_{\rm A}/c$) in the JHU-MHD simulation, for different coarse-graining scales. The contributions of the three force terms are colored as follows: red for $\Theta_\parallel$, blue for $\Theta_\perp$ and magenta for $\boldsymbol{a_E}\cdot\boldsymbol{b}$.  Solid lines are estimated using the first method discussed in the text, dashed lines using the second method. The gray areas delimit the dominant contributions (meaning larger $g$, smaller $t_{\rm int}/g^2$), which are associated to the $\Theta_\parallel$ term.
\label{fig:fill-JHU}}
\end{figure}

In the presence of a broad distribution of values of $Q$ (with $Q$ symbolizing the force contributions as before), the maximum contribution to the diffusion rate at a given $l$ maximizes $Q^2 {\rm d}N/{\rm d}\ln \vert Q\vert$, see Eq.~(\ref{eq:LDPsolp_diff}) above, because $g\propto Q$. In this first exercise, we thus extract this optimal value of $\vert Q\vert$ for each of the three force terms. We then derive the optimal relative energy gain as $g=\vert Q\vert\,l$, see Eq.~(\ref{eq:Rco-evol-p}), and extract the corresponding value of $t_{\rm int}$ as $l/(f_l v)$, where $f_l(>\vert Q\vert)$ represents the filling fraction of regions giving a contribution larger than $\vert Q\vert$ (thus, of either sign),
\begin{equation}
f_l(>\vert Q\vert)\,=\,\frac{1}{N}\int_{\ln\vert Q\vert}^{+\infty}{\rm d}\ln \vert Q'\vert\,\,\left.\frac{{\rm d}N}{{\rm d}\ln \vert Q'\vert}\right\vert_l\,.
\label{eq:fill-frac-Q}
\end{equation}
The results are shown in Fig.~(\ref{fig:fill-JHU}), which suggests that $g \sim 0.4 v_{\rm A}/c$ and $t_{\rm int}/g^2 \sim \mathcal O\left(20\, c^2\ell_{\rm c}/v_{\rm A}^2v\right)$ are reasonable estimates for those parameters. The measured values do not depend strongly on $l$ as long as $l\lesssim \ell_{\rm c}$, suggesting that all scales contribute about equally to the acceleration process. Finally, this exercise also suggests that, at least for the JHU-MHD simulation, the parallel shear contribution $\Theta_\parallel$ seemingly provides the dominant contribution to energy gain.

Our second method for determining those parameters takes into account the possibility of particle trapping inside structures, which happens at values of $Q$ sufficiently large to impart pitch-angle deflections by order of unity. Once inside a structure, the particle gains (or loses) energy in a regular way, as explained in Sec.~\ref{sec:e-drift}. Consequently, trapping may give rise to momentum gains in excess of $Ql$ for some particles. Such effects deserve a careful investigation, which lies beyond the scope of the present paper. Here, we seek a first-order estimate of $t_{\rm int}$ from the structures that are strong enough to give rise to particle trapping, and for such structures, we derive a first-order estimate of $g$ as the minimum (most conservative) value, set by $g= Q l$. The actual gain $g$ is determined by that acquired over the lifetime of the structure, or by the time the particle takes to escape from the structure, whichever happens first. As for any such Fermi process, $g$ is likely distributed as a power law whose index is controlled by the competition between the rate of energy gain $Q$ {\it vs} the rate of escape $\sim 1/l$. At values $Ql$ below unity, this slope is large enough that the average value of $g$ is set by its minimum value, $g \sim Ql$. Physically, this describes a situation in which the particle suffers a pitch-angle deflection of the order of unity through its crossing of the structure, which enhances its energy gain by a factor of unity over the ballistic estimate, but generally not more.

To determine $t_{\rm int}$, we use the distribution of the mirror force term $Q_\mu\equiv \boldsymbol{b}\cdot\boldsymbol{\nabla}\ln B'$ that dominates pitch-angle scattering and search for the filling fraction $f_l(>\vert Q_\mu\vert)$ such that $\vert Q_\mu\vert = 1/l$ (guaranteeing pitch-angle deflection by order unity). We then set $t_{\rm int}=l/f_l(>\vert Q_\mu\vert)$ as before. Finally, we determine, for each of the three force terms $Q$ contributing to particle energization, the energy gain $g$ as $Q l$ where $Q$ is the value associated to the above value of the filling fraction. The results are overplotted in Fig.~\ref{fig:fill-JHU}. They happen to be rather consistent with those found through the previous method, even though the two differ in their spirit. It so happens, however, that the value of $Q_\mu$, as determined through the previous method, is already close to unity, meaning that the structures that maximize the product $Q^2 {\rm d}N/{\rm d}\ln \vert Q\vert$ also tend to trap the particles.

More in-depth studies appear needed to establish a rigorous connection between the properties of the intermittency and the statistics that govern the acceleration rate as a function of scale. Nevertheless, the consistent estimates that emerge from the above exercises suggest that $t_{\rm int}\gtrsim\ell_{\rm c}/v$, $g\lesssim v_{\rm A}/c$  provide reasonable estimates of the parameters governing acceleration in the JHU-MHD simulation. In turn, this suggests an average diffusion coefficient $D_{pp} \propto p^2 (v_{\rm A}/c)^2\, v/\ell_{\rm c}$, as announced earlier, just as it predicts very steep spectra for particle acceleration in this JHU-MHD simulation configuration at a time of several $t_{\rm int} c^2/v_{\rm A}^2$, see Fig.~\ref{fig:sindex}.

\subsection{Further remarks and consequences}\label{sec:e-disc}
The model developed in the previous Sec.~\ref{sec:e-interm}, which describes the momentum gain through a random walk in intermittent structures shares some interesting features with the observations made in dedicated numerical simulations: it agrees on the general scaling of the diffusion coefficient; it predicts a spectral shape similar to that observed, that is a log-normal core slowly drifting in time to larger momenta, with a power law tail; notably so, the spectral index agrees with that measured in those simulations. It would certainly benefit from further detailed analysis aimed at establishing a closer link between the spectral shape and the properties of the turbulence intermittency, as one may notably expect those properties to depend on various external parameters, such as the physics of stirring, the overall magnetization, the amplitude of the turbulent component, the plasma beta parameter etc, {\it e.g.}~\cite{2020PhRvX..10c1021M}. Such studies can involve analytical models, or numerical simulations of particle transport in synthetic intermittent settings, such as undertaken in Refs.~\cite{2020ApJ...895L..14S,2004ApJ...608..540V,2006A&A...449..749T,2016MNRAS.459.3395P,2017PhRvL.119d5101I,2020MNRAS.491.3860S,2020ApJ...895L..14S}. A more detailed analysis of the statistics of interactions between particles and structures in full-blown kinetic (or MHD) simulations of magnetized turbulence would also bring direct constraints and input for the present picture.

We also recall an important prediction of the above model: the (logarithmic) running of the spectral index of the power law tail with momentum. Such a running is not apparent in kinetic numerical simulations, and it should not be apparent given the restricted dynamic range that even the most massive simulations can probe. However, this running clearly deserves further investigation, as it may have a direct impact on phenomenology, for turbulent cascades extending over many orders of magnitude.

To conclude, it proves instructive to recast the above model in the frame of usual quasilinear or Fermi-like treatments of stochastic acceleration and mention a few noteworthy consequences of the impact of intermittency. In contrast to the present microphysical description of particle transport, most treatments and phenomenological applications of stochastic acceleration make use of the Fokker Planck equation to determine the particle distribution function $f(p,t)$. This equation, 
\begin{equation}
\frac{\partial f}{\partial t}\,=\,\frac{1}{p^2}\frac{\partial}{\partial p}\left( p^2 D_{pp}\,\frac{\partial f}{\partial p}\right)\,,
\label{eq:FPstd}
\end{equation}
here written without a drift (other than the noise-induced one) is entirely characterized by the momentum diffusion coefficient $D_{pp}\,\equiv\,\langle\Delta p^2\rangle/(2\Delta t)$. This equation derives from the original Fermi model provided the interaction time between two scattering events does not depend on momentum~\cite{2005PPCF...47B.667D}, corresponding to the scaling $D_{pp}\propto p^2$ observed in numerical simulations~\cite{14Lynn,17Zhdankin,2018ApJ...867L..18Z,2018MNRAS.474.2514Z,18Comisso,2020ApJ...893L...7W,2019ApJ...886..122C,2019PhRvL.122e5101Z}. The same Fokker-Planck equation also derives from the quasilinear transport equation, when rewritten in the diffusion approximation, after averaging over spatial variables~\cite{1975MNRAS.172..557S,1989ApJ...336..243S}.

For $D_{pp}\propto p^2$, the solution for monoenergetic injection at momentum $p_{\rm inj}$ takes a Gaussian shape in $\ln p$,
\begin{equation}
p^3 f\,=\,\left(4\pi \nu_{\rm diff} t\right)^{-1/2}\exp\left\{-\frac{\left[\ln\left(p/p_{\rm inj}\right) - 3\nu_{\rm diff}t\right]^2}{4\nu_{\rm diff}t}\right\}\,
\label{eq:FP-sol}
\end{equation}
with $\nu_{\rm diff}\,=\,D_{pp}/p^2$. The mean of this distribution increases as fast as the variance, indicating substantial drift in momentum, in contrast with the present description; see Sec.~\ref{sec:e-drift}.

Formally, the Fokker-Planck equation represents a truncation to the second order of the Kramers-Moyal expansion of the more fundamental equation that governs the evolution of $f$. This truncation is exact if the corresponding microscopic dynamics can be described by a Langevin equation with Gaussian noise of infinitesimally small coherence time~\cite{1989fpem.book.....R}. It is thus not surprising that this solution fails to reproduce a power law extension in the present context because the intermittency effectively enlarges the coherence time of the random force to the macroscopic timescale $t_{\rm int}$. The above Gaussian shape is rather recovered as the asymptotic limit of the random walk in intermittent turbulence at $t/t_{\rm int}\rightarrow +\infty$, as discussed earlier. Several models have been proposed to generalize the Fokker-Planck equation to incorporate higher-order terms, {\it e.g.} effective resummations~\cite{1990ZhETF..98.1255B,1996ApJ...461L..37B}, fractional transport equations~\cite{2017PhRvL.119d5101I}, or more pragmatically, a Fokker-Planck equation with a momentum-dependent advection coefficient, tuned in such a way as to reproduce the results of kinetic simulations~\cite{2020ApJ...893L...7W}.

The role played by intermittency may have other, direct implications for phenomenology. For one, strong intermittency is associated to inhomogeneity, which directly implies nonuniformity in space and time of the radiation emitted by the accelerated particles~\cite{2020MNRAS.493..603Z,2021ApJ...908...71Z,2020arXiv201203043N}. It can also impact the spatial transport of particles, just as it impacts their transport in momentum space. In particular, if the interaction length between two scattering events is $\sim \ell_{\rm c}$, as suggested by the analysis of the previous section, then the fraction of unscattered particles must remain of the order of unity on length scales $\ell_{\rm c}$. This implies that significant anisotropies can be maintained on long length scales, with obvious consequences for cosmic-ray physics.

\section{Summary and conclusions}\label{sec:conclusion}
This paper has discussed the physics of particle acceleration in strong MHD turbulence characterized by an ensemble of structures, rather than by the linear superposition of random waves of quasilinear theory. To this effect, it has made use of a novel formalism, applicable from the subrelativistic to relativistic limits, that describes the history of the particle in a mixed phase space, with spatial variables expressed in the lab-frame, and momenta in the \Rco~frame that moves at velocity $\boldsymbol{\beta_E}=\boldsymbol{E}\times\boldsymbol{B}/B^2$ (in units of $c$; its four-velocity is written $\boldsymbol{u_E}$), where the motional electric field vanishes. All energy gains or losses can then be expressed as the influence of inertial forces that derive from the noninertial nature of that frame \Rco. These energy gains/losses are thus directly related to the gradient of the velocity flow of the magnetic field lines, in particular their acceleration, their compression and their shear. This allows us to connect in a direct manner the physics of particle acceleration to the description of the turbulent bath in terms of velocity structures.  To quantify the terms of energy gain/loss, we have assumed that the particle undergoes local gyromotion around the perturbed field lines and retained the lowest order contribution in $r_{\rm g}/\ell_{\rm c}$ (ratio of particle gyroradius to coherence length scale). The main results can be summarized as follows:

\begin{enumerate}

    \item Three terms contribute to particle acceleration: the effective gravity projected along the field line ($\boldsymbol{a_E}\cdot\boldsymbol{b}$), associated to the acceleration field $\boldsymbol{a_E}$ of $\boldsymbol{u_E}$, the shear of $\boldsymbol{u_E}$ along the field line ($\Theta_\parallel$), and its compression in the transverse plane ($\Theta_\perp$); the last two dominate over the first in subrelativistic turbulence. In ideal MHD, and to the present order of approximation, those forces capture the classical drifts of guiding center models. The present formalism is amenable to extensions that can capture terms of higher order in $r_{\rm g}/\ell_{\rm c}$ or nonideal contributions to Ohm's law.

    \item The present model is closer in spirit to the original Fermi scenario of stochastic particle acceleration than to quasilinear theory (QLT). It differs from this original Fermi scenario in one crucial aspect: in each region of nonzero gradient $\Theta_\parallel$ or $\Theta_\perp$, the particle gains (or loose) energy in a regular (nonstochastic) manner that depends on the sign of the gradient, not whether the collision with that region is head-on or tail-on. The acceleration process becomes diffusive once the particle has encountered several uncorrelated energizing sites. 

    \item On average, the particles suffer net energy gain or loss, whose amount is determined by the volume average of those gradients (weighted by the particle distribution). This mean energy gain can be directly related to the net heating of the plasma through large-scale motions. This also differs from the usual quasilinear treatment, which implicitly assumes an infinite reservoir of energy for particle energization, and which predicts similar amounts of net and stochastic energy gains.

    \item As demonstrated by measurements carried out in the JHU-MHD $1024^3$ numerical simulation of incompressible MHD, the acceleration sites are localized in isolated regions that occupy a small filling fraction of the overall turbulent volume, unlike QLT, which assumes a homogeneous space-filling background of waves. Furthermore, those acceleration sites are preferentially found in regions of strong gradients of the magnetic energy density, whereas QLT predicts an acceleration rate that scales with the turbulent magnetic energy density.

    \item The circumscription of accelerating sites to sparse regions is a hallmark of intermittency. Correspondingly, the statistics of the velocity gradients exhibit highly non-Gaussian features characterized by power law tails leading to large excursions, all the more so as the coarse-graining scale decreases.

    \item Intermittency plays a key role in shaping the distribution function of the accelerated particles. The spectrum has been derived here in the frame of a microscopic model describing the transport of particles in a bath of intermittent acceleration sites. This spectrum exhibits a Gaussian core that broadens in time, as in a diffusive process, together with a (near-)power law at large momenta, which  hardens with time. Overall, the spectral shape will thus differ from that evaluated assuming a homogeneous turbulence, which opens interesting prospects for a rich phenomenology.

    \item In such a situation, the average diffusion coefficient characterizes the broadening of the Gaussian core, but not the power law tail. The reason why the standard Fokker-Planck treatment fails here is because the intermittency introduces a macroscopic timescale $t_{\rm int}$, defined as the typical interaction time with an accelerating site. In the JHU-MHD simulation, $t_{\rm int}\sim \ell_{\rm c}/v$ ($v$ particle velocity). The corresponding diffusion coefficient $D_{pp} \propto p^2 \langle \delta u_E^2\rangle v/\ell_{\rm c}$ agrees qualitatively well with that measured in test particle simulations of subrelativistic, weakly sub-Alfv\'enic MHD turbulence~\cite{14Lynn}, as well as that measured in fully kinetic simulations of relativistic turbulence~\cite{17Zhdankin,2018ApJ...867L..18Z,2018MNRAS.474.2514Z,18Comisso,2020ApJ...893L...7W,2019ApJ...886..122C,2019PhRvL.122e5101Z}.

    \item The particles that populate the nonthermal power law are those that have been able to interact with sufficiently many sites in a given time. At time $t \sim t_{\rm int}/g^2$, with $g$ the characteristic relative energy gain per interaction with an accelerating site, the spectral index $s$ (${\rm d}N/{\rm d}p\propto p^{-s}$) is predicted to be of the order of $3$ for $g\sim \mathcal O(1)$. This agrees rather well with the aforementioned kinetic numerical simulations of relativistic turbulence. Smaller values of $g$ (as expected in subrelativistic turbulence) lead to softer spectra. 

\end{enumerate}

As discussed in Sec.~\ref{sec:structures}, the energization in regions of large shear of the velocity field of magnetic field lines is closely related to the sources and sinks of magnetic energy in the MHD description, and more generally with intermittent structures characterized by large gradients. This should be put in perspective with numerous observations and simulations that have revealed that plasma heating in turbulent environments is indeed associated with intermittent structures~\cite{2009PhRvL.103g5006K,2011ApJ...727L..11O,2012PhRvL.108z1102O,2019E&SS....6..656B,2012PhRvL.109s5001W,2013PhPl...20a2303K,2014PhRvL.112u5002O}, 
albeit generally interpreted as the consequence of reconnection.

Those two pictures enjoy intriguing connections. For one, reconnecting currents sheets are associated with violent bulk plasma motions that can be described in an MHD approximation over most of the volume, outside the diffusion regions. In such environments, particle energization is dominated by Fermi-type interactions, at least in the absence of a strong guide field, and all the more so at high energies, {\it e.g.}~\cite{2016PhPl...23l0704D,2020PhPl...27h0501G}. More precisely, the perpendicular electric field that governs those Fermi-type processes can be related to the shearing and compressive motions of the flow, as in the present formalism~\cite{2018ApJ...855...80L}. Zooming out to larger scales, particles could also be accelerated through repeated encounters with the exhausts of many, randomly oriented reconnection regions~\cite{12Hoshino}, or more generally with the nonuniform stirring motions around  reconnection regions, if reconnection takes place in a turbulent fashion on all scales~\cite{1999ApJ...517..700L}. Therefore, what truly governs plasma heating in the vicinity of intermittent structures deserves closer scrutiny, from observational, theoretical and numerical viewpoints. 

One merit of the present model is that it can be directly tested against observations or numerical simulations by comparing the observed amount of particle energization with the expected contributions for the inertial, the parallel shear and the transverse compression terms. To calculate those expected contributions, it suffices to measure the statistics of the gradients of $\boldsymbol{u_E}$ and use Eq.~(\ref{eq:Rco-evol}). In that regard, Fig.~\ref{fig:Bmap} provides a direct tomography of the acceleration regions in the JHU-MHD simulation. 

This contrasts somewhat with the quasilinear view of energization through wave particle resonant interactions, which are difficult to observe; see however Ref.~\cite{2017JPlPh..83d5301K} for recent progress in that direction. For that reason, one commonly tests models of particle acceleration (or transport) through the comparison of the predicted diffusion coefficient with that measured in  simulations. However, this does not suffice, because various theoretical models can lead to the same overall scaling with particle momentum, and furthermore, the diffusion coefficient itself does not fully characterize the acceleration process, as discussed above. 

The role of turbulence intermittency regarding particle transport, a novel aspect of the present model, deserves further investigation. One of its most obvious signatures to look for, in observations or numerical simulations, is inhomogeneity in the rates of particle diffusion; such inhomogeneity has actually been observed in Refs.~\cite{2020ApJ...894..136T,2020arXiv201203043N}. Intermittency is likely to entail important consequences for phenomenological applications in space plasma physics. For instance, it may lead to significant anisotropies in phase space on scales comparable to the coherence length, for the same reason that it affects the momentum distribution function. One should thus aim at establishing a closer relationship between the properties of the turbulence (nature of the forcing, magnetization, amplitude etc.) and the phenomenology of particle transport and acceleration. This may involve test-particle propagation in synthetic turbulence as well as full numerical simulations of turbulence with advanced diagnostics.

\begin{acknowledgments}
The possibility to use the resources of the JH Turbulence Database (JHTDB), which is developed as an open resource by the Johns Hopkins University, under the sponsorship of the National Science Foundation, as well as the tools made available for public use, is gratefully acknowledged. This work has been supported in part by the Sorbonne Universit\'e DIWINE Emergence-2019 program. It is a pleasure to thank V. Bresci, A. Bykov, L. Comisso, C. Demidem, L. Gremillet, M. Malkov, L. Sironi, and L. Vlahos for insightful discussions.
\end{acknowledgments}

\appendix

\section{PARTICLE ACCELERATION IN \Rco}\label{sec:appA}
The following conventions are used. Greek indices refer to spacetime indices in the laboratory frame, while latin indices $(a,\,b,\,\ldots)$ are spacetime indices in the frame \Rco~in which the comoving electric field vanishes. Correspondingly, tensors indexed with greek indices are understood to be evaluated in the lab frame, while those indexed with latin indices are expressed in the \Rco~frame. Where confusion can arise, a prime is used to label comoving frame quantities. Hence $\boldsymbol{B'}$ denotes the magnetic field in the comoving frame, while $\boldsymbol{B}$ (as the electric field $\boldsymbol{E}$) that in the lab frame. These two are related through the usual (instantaneous) Lorentz transforms,  
\begin{align}
&\boldsymbol{B'}\,=\,\gamma_E\,\boldsymbol{B}-\boldsymbol{u_E}\times\boldsymbol{E}\,,\quad\quad
\boldsymbol{B}\,=\,\gamma_E\boldsymbol{B'}\,,\nonumber\\
&\boldsymbol{E}\,=\,-\boldsymbol{u_E}\times\boldsymbol{B'}\,,
\label{eq:LtransformEB}
\end{align}
with $B'=\sqrt{B^2-E^2}$. The quantities $\gamma_E=B/B'$ and $\boldsymbol{\beta_E}=\boldsymbol{E}\times\boldsymbol{B}/B^2$ represent the Lorentz factor and three-velocity of \Rco~in units of $c$; its four-velocity is written $\boldsymbol{u_E}=\boldsymbol{\beta_E}/\sqrt{1-{\beta_E}^2}$ in the lab frame. In the \Rco~frame, ${u_E}^a\,=\,\left(1,\boldsymbol{0}\right)$.

Following \cite{2019PhRvD..99h3006L}, the four-velocity of the particle in the locally inertial \Rco~frame, ${u'}^a$, evolves according to
\begin{equation}
\frac{1}{c}\frac{{\rm d} {u'}^a}{{\rm d}\tau}\,=\,\frac{q}{m}\,
{F^a}\!_b\, {u'}^b\,-\,
\widehat\Gamma^a_{bc}\,{u'}^b\,{u'}^c\,,
\label{eq:dyn}
\end{equation}
with $F^{ab}$ the Maxwell field strength tensor in \Rco~(thus, purely magnetic), and $\widehat\Gamma^a_{b\,c}$ the connection that captures the inertial forces,
\begin{equation}
\widehat \Gamma^a_{bc}\,=\,- {{\mathsf e}^\beta}_b\,{{\mathsf e}^\gamma}_c\,\frac{\partial}{\partial x^\gamma}{{\mathsf e}^{a}}_\beta\,,
\label{eq:conn1}
\end{equation}
here expressed as a function of the tetrad ${{\mathsf e}^\beta}_b$ (and its inverse) that connects the lab to the \Rco~frame, see {\it e.g.}~\cite{1985ApJ...296..319W,1989ApJ...340.1112W}. Note that the derivative is taken with respect to proper time $\tau$ which parametrizes the particle worldline. 

The choice of this tetrad is not unique and, in the present case, it proves useful to define it as follows~\cite{1978JPhA...11.1069F,1987PhDT.......197B}
\begin{align}
\left\{{\mathsf{e}^\mu}_a,\,\,a=0,3\right\}&\,=\,\left\{{u_E}^\mu,\,b^\mu,\,{e_2}^\mu,\,{e_3}^\mu\right\}\,\nonumber\\
\left\{{\mathsf{e}^a}_\mu,\,\,a=0,3\right\}&\,=\,\left\{-{u_E}_\mu,\,b_\mu,\,{e_2}_\mu,\,{e_3}_\mu\right\}\,
\label{eq:tetrad}
\end{align}
with ${u_E}^\mu$ the (timelike) four-velocity of \Rco~in the lab frame. The spacelike four-vector $b^\mu$ is the unit vector defined by $b^\mu = B^\mu/\sqrt{B^\alpha B_\alpha}$, with 
\begin{equation}
B^\mu\,=\,-^{\star}{F^\mu}\!_{\nu}\,{u_E}^\nu\, ,
\label{eq:Bmagvec}
\end{equation}
the magnetic four-vector, which represents the magnetic field measured by an observer moving at velocity ${u_E}^\nu$: $B^\mu\,=\,\left\{0,\boldsymbol{B'}\right\}$ in the lab-frame.  The dual field strength tensor is written $^\star F^{\mu\nu}\,=\,\frac{1}{2}\epsilon^{\mu\nu\alpha\beta}F_{\alpha\beta}$, with $\epsilon^{0123}=-1$. The last two spacelike vectors ${e_2}^\mu$ and ${e_3}^\mu$ span the plane orthogonal to ${u_E}^\mu$ and $b^\mu$. Their exact definition is not important in what follows. They can be defined through linear combinations of the eigenvectors of ${F^a}\!_b$ with eigenvalues $+iB'$ and $-i B'$.

This choice of tetrad tacitly implies a rotation of the three spatial axes with respect to the lab frame. In particular, the magnetic four-vector in \Rco~is written $B^a\,=\,{{\mathsf e}^a}_\mu B^\mu\,=\,\left\{0,B',0,0\right\}$. 

In the \Rco~frame, the four-velocity of the particle is written $u^a\,=\,\left\{\gamma',\,u_\parallel',\, u'_{2},\,u'_{3}\right\}$. Noting that the Lorentz force in Eq.~(\ref{eq:dyn}) contributes neither to the evolution of $\gamma'$ nor to that of $u_\parallel'$, we obtain
\begin{align}
\frac{1}{c}\frac{{\rm d}\gamma'}{{\rm d}\tau}&\,=\,-\gamma'u_\parallel'\,b^\beta {u_E}^\gamma \partial_\gamma {u_E}_\beta - {u_\parallel'}^2\,b^\beta b^\gamma \partial_\gamma {u_E}_\beta
\nonumber\\&\quad - {u_{2}'}^2{e_2}^\beta {e_2}^\gamma \partial_\gamma {u_E}_\beta
- {u_{3}'}^2{e_3}^\beta {e_3}^\gamma \partial_\gamma {u_E}_\beta\,,\nonumber\\
\frac{1}{c}\frac{{\rm d}u_\parallel'}{{\rm d}\tau}&\,=\,{\gamma'}^2\, {u_E}^\beta {u_E}^\gamma \partial_\gamma b_\beta + \gamma'u_\parallel'\,{u_E}^\beta b^\gamma \partial_\gamma b_\beta
\nonumber\\&\quad + {u_{2}'}^2{e_2}^\beta {e_2}^\gamma \partial_\gamma b_\beta
+ {u_{3}'}^2{e_3}^\beta {e_3}^\gamma \partial_\gamma b_\beta\,.
\label{eq:dynRco1}
\end{align}
Crossed terms containing only one power of $u_{2}'$ or $u_{3}'$ have been neglected, for reasons discussed further below. 

The force terms in Eq.~(\ref{eq:dynRco1}) find a clear meaning after being rewritten in terms of derivatives of the velocity field. To do this, decompose $\partial_\beta{{u_E}^\alpha}$ as
\begin{equation}
\partial_\beta{u_E}^\alpha\,=\,{{\sigma_E}^{\alpha}}\!_{\beta}\,+\,{{\omega_E}^\alpha}\!_\beta
\,+\, \frac{1}{3}\Theta_E\,{{h_E}^\alpha}\!_\beta\,+\,{{a_E}^\alpha}\!_\beta\,,
\label{eq:uEdecompos}
\end{equation}
in terms of ${{\sigma_E}^{\alpha}}\!_{\beta}$ the shear tensor, ${{\omega_E}^\alpha}\!_\beta$ the vorticity tensor, $\Theta_E$ the expansion scalar, ${{h_E}^\alpha}\!_\beta$ the orthogonal projector and ${{a_E}^\alpha}\!_\beta$ the acceleration tensor, defined through
\begin{align}
{h_E}^{\alpha\beta}&\,=\,\eta^{\alpha\beta}\,+\,{u_E}^\alpha {u_E}^\beta\,,\nonumber\\
{\sigma_E}^{\alpha\beta}&\,=\,\frac{1}{2}{h_E}^{\alpha\mu}{h_E}^{\beta\nu}\left(\partial_\nu {u_E}_\mu + 
\partial_\mu {u_E}_\nu\right)-\frac{1}{3}\Theta_E\,{h_E}^{\alpha\beta}\,,\nonumber\\
{\omega_E}^{\alpha\beta}&\,=\,\frac{1}{2}{h_E}^{\alpha\mu}{h_E}^{\beta\nu}\left(\partial_\nu {u_E}_\mu - 
\partial_\mu {u_E}_\nu\right)\,,\nonumber\\
\Theta_E&\,=\,\partial_\alpha {u_E}^\alpha\,,\nonumber\\
{a_E}^{\alpha\beta}&\,=\, -{u_E}^\beta {u_E}^\mu\partial_\mu {u_E}^\alpha\,.
\label{eq:uEdecompos2}
\end{align}
The acceleration four-vector is ${a_E}^\alpha= {u_E}_\beta\,{a_E}^{\alpha\beta}={u_E}^\mu\partial_\mu {u_E}^\alpha$. Given that the magnetic field vector defines a preferred direction, we split $\Theta_E\,=\,\Theta_{\parallel}+\Theta_{\perp}$ with
\begin{align}
\Theta_{\parallel}&\,=\,b^\alpha b^\beta \partial_\alpha {u_E}_\beta\,,\nonumber\\
\Theta_{\perp}&\,=\,\left(\eta^{\alpha\beta}  - b^\alpha b^\beta\right) \partial_\alpha {u_E}_\beta\,,
\label{eq:thetaperp}
\end{align}
which characterize the shear along and the compression perpendicular to the magnetic field line. 

This eventually leads to the main equation,
\begin{align}
\frac{1}{c}\frac{{\rm d}\gamma'}{{\rm d}\tau}&\,=\,-\gamma'u_\parallel'\,\boldsymbol{a_E}\cdot\boldsymbol{b} - {u_\parallel'}^2\,\Theta_\parallel- \frac{1}{2}{u_{\perp}'}^2\Theta_\perp\,,\nonumber\\
\frac{1}{c}\frac{{\rm d}u_\parallel'}{{\rm d}\tau}&\,=\,-{\gamma'}^2\, \,\boldsymbol{a_E}\cdot\boldsymbol{b} - \gamma'u_\parallel'\,\Theta_\parallel - \,\frac{1}{2}{u_{\perp}'}^2\boldsymbol{b}\cdot\boldsymbol{\nabla}\ln\,B'.
\label{eq:Rco-evol-app}
\end{align}

The first two terms on the rhs of each equation derive directly from the decomposition of $\partial_\gamma {u_E}_\beta$ and $\partial_\gamma b_\beta$. The third are obtained as averages over a gyroperiod, assuming that on such short timescales, the gradients do not change significantly. Then, $\langle u_2'\rangle=\langle u_3'\rangle =\langle u_2' u_3'\rangle=0$ but $\langle {u_{2}'}^2\rangle=\langle {u_{3}'}^2\rangle = {u_\perp'}^2/2$. All terms containing only one power of $u_{2}'$ or $u_{3}'$ have already been excluded from Eq.~(\ref{eq:dynRco1}). The decomposition of the identity, $\eta^{\alpha\beta} = -{u_E}^\alpha{u_E}^\beta + b^\alpha b^\beta + {e_2}^\alpha {e_2}^\beta + {e_3}^\alpha {e_3}^\beta$, combined with ${u_E}_\alpha\partial_\beta{u_E}^\alpha=0$ then leads to the last term in the equation for $\gamma'$. 

Defining ${h_\perp}^{\alpha\beta}\equiv {e_2}^\alpha {e_2}^\beta + {e_3}^\alpha {e_3}^\beta$, the corresponding term $\propto b^\alpha \partial_\alpha B'$ in the equation for $u_\parallel'$ is obtained through
\begin{align}
{h_\perp}^{\beta\gamma}\partial_\gamma b_\beta&\,=\,
\frac{1}{B'}{h_\perp}^{\beta\gamma}\partial_\gamma B_\beta\nonumber\\
&\,=\,\frac{1}{B'}\left(\eta^{\beta\gamma}+{u_E}^\beta{u_E}^\gamma-b^\beta b^\gamma\right)\partial_\gamma B_\beta\nonumber\\
&\,=\,\frac{1}{B'}D^\alpha B_\alpha -  \frac{1}{B'}b^\alpha\partial_\alpha B'.
\label{eq:dynRco3}
\end{align}
The differential operator $D^\alpha$ denotes the orthogonally projected derivative relatively to the timelike direction set by ${u_E}^\alpha$, see~\cite{2005CQGra..22..393T,2007PhR...449..131B},
\begin{align}
D^\alpha B_\alpha \,\equiv \, {{h_E}^\alpha}\!_\mu{{h_E}^\nu}\!_\alpha \partial_\nu B^\mu
\,=\, {{h_E}^\nu}\!_\mu \partial_\nu B^\mu\,.
\label{eq:orthcovder}
\end{align}
and, in the MHD approximation, $D^\alpha B_\alpha = 0$ identically; that latter equation generalizes $\boldsymbol{\nabla}\cdot\boldsymbol{B}=0$ to accelerated frames. 

So far, everything has been written in terms of partial, noncovariant derivatives, but general covariantization is straightforward.

\subsection{Connection to the guiding center approximation}
The guiding center approximation relies on the conservation of the magnetic moment,
\begin{equation}
M\,\equiv\,\frac{{u_\perp'}^2}{2B'}\,,
\label{eq:def-M}
\end{equation}
and it is best described (to lowest order) by the covariant relativistic Hamiltonian~\cite{1981PhFl...24.1730L},
\begin{equation}
H\,=\,\frac{1}{2}m\overline u_\mu\,\overline u^\mu\,+\,m M B'\,,
\label{eq:H0gc}
\end{equation}
in terms of $\overline u^\mu$ the four-velocity of the guiding center, defined as the particle four-velocity deprived of gyromotion,
\begin{equation}
\overline u^\mu\,=\,u^\mu - u_{2}' {e_2}^\mu- u_{3}' {e_3}^\mu\,=\,\gamma'{u_E}^\mu + {u_\parallel'}b^\mu\,.
\label{eq:def-ubar}
\end{equation}
The equations of motion can be derived using the Poisson brackets,
\begin{align}
\left\{\overline x^\alpha,\,  \overline u^\beta\right\}&\,=\,\frac{1}{m}\left(-{u_E}^\alpha {u_E}^\beta\,+\,b^\alpha b^\beta\right),\,
\label{eq:Poisson-brack}
\end{align}
which gives to lowest order,
\begin{align}
\frac{1}{c}\frac{{\rm d}{\overline x}^\alpha}{\rm d \tau}&\,=\,\gamma' {u_E}^\alpha\,+\, u_\parallel' b^\alpha\,,\nonumber\\
\frac{1}{c}\frac{{\rm d}{\overline u}^\beta}{\rm d \tau}&\,=\,-M\left(-{u_E}^\beta {u_E}^\gamma\,+\,b^\beta b^\gamma\right)\,\partial_\gamma B'\,.
\label{eq:gc-position}
\end{align}
The latter equation can be rewritten for $\gamma'$ and $u_\parallel'$, 
\begin{align}
\frac{{\rm d} \gamma'}{{\rm d}\tau}& \,=\,u_\parallel'\,{u_E}^{\alpha}\frac{{\rm d} b_\alpha}{{\rm d}\tau} + M c\,{u_E}^\alpha \partial_\alpha B' \,,\nonumber\\
\frac{{\rm d}u_\parallel'}{{\rm d}\tau}& \,=\,-\gamma'\,b_\alpha\frac{{\rm d} {u_E}^\alpha}{{\rm d}\tau} - M c\,b^\alpha \partial_\alpha B' \,.
\label{eq:gc-momenta-evol}
\end{align}
Combining both equations of Eq.~(\ref{eq:gc-momenta-evol}) with the four-velocity normalization $\overline u'_a \overline u'^a+2MB'=-1$ leads to ${\rm d}M/{\rm d}\tau = 0$, expressing conservation of the adiabatic moment.

We can establish a connection between the two formalisms and rewrite Eq.~(\ref{eq:gc-momenta-evol}) in the form of Eq.~(\ref{eq:Rco-evol-app}), as follows. First note that the inertial contribution, {\it viz.} the first term on the rhs of Eq.~(\ref{eq:gc-momenta-evol}) breaks down into the first two terms on the rhs of Eq.~(\ref{eq:Rco-evol-app}), noting that, in the guiding center approximation,
\begin{align}
\frac{1}{c}\frac{{\rm d}}{{\rm d}\tau}&\,\equiv\,\langle\gamma\rangle\frac{1}{c}\frac{\partial}{\partial t}\,+\, u_\parallel \boldsymbol{b}\cdot\boldsymbol{\nabla}\,+\,
\langle\gamma\rangle\boldsymbol{\beta_E}\cdot\boldsymbol{\nabla}\,,\nonumber\\
&\,\equiv\,\gamma'{u_E}^\alpha\partial_\alpha\,+\,u_\parallel'b^\alpha\partial_\alpha\,,
\label{eq:d2dtau}
\end{align}
because
\begin{align}
{u_E}^\alpha\partial_\alpha& \,=\,\gamma_Ec^{-1}\partial_t\,+\,\boldsymbol{u_E}\cdot\boldsymbol{\nabla}\,=\,c^{-1}\partial_{t'}\,,\nonumber\\
b^\alpha\partial_\alpha&\,=\, \boldsymbol{b}\cdot\boldsymbol{\nabla}\,=\,\boldsymbol{b'}\cdot\boldsymbol{\nabla'}\,.
\label{eq:partderiv}
\end{align}

Those two terms on the rhs of Eq.~(\ref{eq:d2dtau}) lead to the the effective gravity force projected along the field line direction (${a_E}^\alpha\,b_\alpha$) and the shear acceleration or compression along the field line direction ($\Theta_\parallel$). The latter characterizes the curvature drift part of the energy change. This is because the curvature drift velocity scales as $\boldsymbol{v_{\rm d,\,curv}}\propto \boldsymbol{B}\times\left(\boldsymbol{b}\cdot\boldsymbol{\nabla}\boldsymbol{b}\right)$, so that $\boldsymbol{E}\cdot\boldsymbol{v_{\rm d,\,curv}} \propto \left(\boldsymbol{E}\times \boldsymbol{B}\right)\cdot\left(\boldsymbol{\nabla}\boldsymbol{b}\right)\propto \boldsymbol{b}\cdot\left(\boldsymbol{b}\cdot\boldsymbol{\nabla}\boldsymbol{u_E}\right)$, given that $\boldsymbol{b}\cdot\boldsymbol{u_E}=0$.

The third terms on the rhs of Eqs.~(\ref{eq:Rco-evol-app}) and (\ref{eq:gc-momenta-evol}) match one another in the expression for $u_\parallel'$; they describe the grad-B force.  In the equation for $\gamma'$, this grad-B term can also be rewritten in terms of derivatives of the velocity field through the projection of the advection equation on $b_\beta$,
\begin{align}
& b_\beta\partial_\alpha\left({u_E}^\alpha B^\beta - {u_E}^\beta B^\alpha\right)\,=\,0\nonumber\\
\Rightarrow \quad&{u_E}^\alpha\partial_\alpha B' \,=\,-B'\left(\eta^{\alpha\beta}-b^\alpha b^\beta\right)\partial_\alpha{u_E}_{\beta}\,\nonumber\\
& \quad\quad\quad\quad\, =-\Theta_\perp B'\,.
\label{eq:advect}
\end{align}
Hence, the grad-B term appears here under a variety of forms: it can be depicted as a betatron process, as in Eq.~(\ref{eq:V60drift}), as a mirror force, or as the result of compression in the plane transverse to the field lines.

Finally, one can check that these equations match the guiding center equations of motion of Vandervoort and Northrop~\cite{1960AnPhy..10..401V,1961AnPhy..15...79N},
\begin{equation}
\frac{{\rm d}\langle\gamma\rangle}{{\rm d}\tau}\,=\,
\frac{q}{mc^2}\boldsymbol{E}\cdot\boldsymbol{v_{\rm d}}\,+\,M\frac{\partial}{\partial t}B'\,,
\label{eq:V60drift}
\end{equation}
where $\langle\gamma\rangle = \gamma_E\gamma'$ corresponds to the Lorentz force of the particle in the lab-frame, averaged over a gyroperiod, with drift velocity,
\begin{align}
\boldsymbol{v_{\rm d}}\,=\,\frac{m c^2}{q}\biggl\{& \frac{u_\parallel'}{{B'}^2}\boldsymbol{B}\times\frac{{\rm d}\boldsymbol{b}}{{\rm d}\tau} + \frac{\langle\gamma\rangle}{{B'}^2}\boldsymbol{B}\times\frac{{\rm d}\boldsymbol{\beta_E}}{{\rm d}\tau}\nonumber\\
&\quad\,+\,
\frac{Mc}{{B'}^2}\boldsymbol{B}\times\boldsymbol{\nabla}B'\,+\,
\frac{M}{{B'}^2}\boldsymbol{E}\frac{\partial B'}{\partial t}\biggr\}\,.
\label{eq:V60drift2}
\end{align}
Rewriting Eq.~(\ref{eq:V60drift}) above for $\gamma'$ instead of $\langle\gamma\rangle$, it can be shown that the inertial drift contribution reproduces the first term of Eq.~(\ref{eq:gc-momenta-evol}), while the polarization drift vanishes in the proper frame, and the grad-B drift terms can be reorganized as the second term of Eq.~(\ref{eq:gc-momenta-evol}). Similar operations can be conducted to recover the evolution of $u_\parallel'$.\bigskip

\section{MOMENTUM DIFFUSION IN INTERMITTENT TURBULENCE}\label{sec:appB}
The random walk described in Sec.~\ref{sec:e-interm} can be modeled as the adimensioned process
\begin{equation}
\hat y\,=\,\frac{1}{\mathcal N}\sum_{i=1}^{\mathcal N}\,\tilde x_i\,,
\label{eq:rw-ldp}
\end{equation}
with the short-hand notations: $y = \ln (p/p_{\rm inj})/\left({\mathcal N}g\right)$, $g$  the mean energy gain, ${\mathcal N}=t/\tau$ the number of jumps, each taking time $\tau$; $\tilde x_i$ a random variable that takes value $+1$ with (energy gain) probability $f_+$, $-1$ with (energy loss) probability $f_-$ and $0$ with probability $f_{\emptyset}\equiv1-f_--f_+$. 

The probability distribution function $\phi(\hat y)$ can be approximated using the techniques of large deviation theory~\cite{2009PhR...478....1T}, as
\begin{equation}
\phi(\hat y)\,\underset{{\mathcal N}\gg1}{\sim}\,\exp\left[-{\mathcal N} I(\hat y)\right]\,,
\label{eq:ldp-0}
\end{equation}
where 
\begin{equation}
I(\hat y) = \underset{\upsilon\in\mathbb{R}}{\rm sup}\left\{\upsilon \hat y - \psi(\upsilon)\right\}\,,
\label{eq:ldp-1}
\end{equation}
the function $\psi(\upsilon)$ being defined through the cumulant generating function of the random variable $\tilde x$,
\begin{align}
\psi(\upsilon) &\,=\, \ln\left\{\mathbb{E}\left[e^{\upsilon \tilde x}\right]\right\}\nonumber\\
&\,=\, \ln\left(f_- e^{-\upsilon} + f_+ e^{+\upsilon} + f_{\emptyset}\right)\,.
\label{eq:ldp-2}
\end{align}
This gives
\begin{align}
I(\hat y)\,=\,&\hat y\,\ln\left[\frac{f_{\emptyset} \hat y + \sqrt{f_{\emptyset}^2{\hat y}^2 + 4f_-f_+(1-{\hat y}^2)}}{2f_+(1-{\hat y})}\right]\nonumber\\
&\quad - \ln\left[\frac{f_{\emptyset} + \sqrt{f_{\emptyset}^2{\hat y}^2 + 4f_-f_+(1-{\hat y}^2)}}{(1-{\hat y}^ 2)}\right]\,.
\label{eq:ldp-3}
\end{align}

The mean value for $\hat y$ is $\hat y_m\equiv \langle\hat y\rangle = f_+-f_-$, as is obvious from Eq.~(\ref{eq:rw-ldp}). At $\hat y_m$, we have
\begin{align}
I(\hat y_m)&\,=\,0\,,\quad\quad 
I'(\hat y_m)\,=\,0\,,\nonumber\\
I''(\hat y_m)&\,=\, \frac{1}{f_-+f_+ - (f_+-f_-)^2}\,.
\label{eq:ldp-4}
\end{align} 
Consequently, for $\hat y \sim \hat y_m$, corresponding to $\ln (p/p_0) \sim {\mathcal N}g(f_+-f_-)$, the distribution is nearly Gaussian,
\begin{equation}
\phi(\hat y)\,\underset{{\mathcal N}\rightarrow+\infty}{\sim}\,
\exp\left[-\frac{\mathcal N}{2}I''(\hat y_m)\left(\hat y - \hat y_m\right)^2\right]\,,
\label{eq:ldp-5}
\end{equation}
but, on intermediate timescales $\mathcal N < +\infty$, it exhibits a non-Gaussian tail at large values of $\hat y$.

\bibliography{refs}

\end{document}